\documentclass{ws-ijgmmp}

\usepackage[svgnames]{xcolor}
\usepackage[colorlinks]{hyperref}
\usepackage{float}
\usepackage{soul}
\usepackage{accents}

\usepackage{threeparttable}

\usepackage{tikz}
\usepackage{cite}
\usepackage{tikzsymbols}
\usepackage[utf8]{inputenc} 
\usepackage{times}
\usepackage{mathtools}
\usepackage{footnote}
\makesavenoteenv{table}
\usetikzlibrary{arrows,shapes}
\usetikzlibrary{matrix}
\usetikzlibrary{shadows}
\usetikzlibrary{positioning}

\tikzset{
    >=stealth',
    pil/.style={
           ->,
           thick,
           shorten <=2pt,
           shorten >=2pt,}
}

\newcommand{\metric}{g}
\newcommand{\tetrad}{\theta}
\newcommand{\cotetrad}{e}

\newcommand{\spinconnection}{\omega}
\newcommand{\dettetrad}{\theta}
\newcommand{\Lorentz}{\Lambda}
\newcommand{\Ricci}{R}
\newcommand{\lapse}{\alpha}
\newcommand{\shift}{\beta}
\newcommand{\inducedmetric}{\gamma}
\newcommand{\normalvector}{n}
\newcommand{\momenta}{\pi}

\newcommand{\dd}{\mathrm{d}}
\newcommand{\intprod}{\mathrel{\reflectbox{\rotatebox[origin=c]{180}{$\neg$}}}}
\newcommand{\vek}[1]{\prescript{\mathcal{V}}{}{#1}}
\newcommand{\asy}[1]{\prescript{\mathcal{A}}{}{#1}}
\newcommand{\sym}[1]{\prescript{\mathcal{S}}{}{#1}}
\newcommand{\trc}[1]{\prescript{\mathcal{T}}{}{#1}}
\newcommand{\pr}{\Xi}
\newcommand{\tngr}{\mathbb{T}_{\textrm{NGR}}}
\newcommand{\kernelcomp}{l}
\newcommand{\lc}[1]{\mathring{#1}}
\begin{document}

	\markboth{D. Blixt, M.J. Guzm\'an., M. Hohmann, and C. Pfeifer}
	{Review of the Hamiltonian analysis in teleparallel gravity}

	\catchline{}{}{}{}{}

	\title{Review of the Hamiltonian analysis in teleparallel gravity
	}

	\author{Daniel Blixt}
	\address{Laboratory of Theoretical Physics, Institute of Physics, University of Tartu, W. Ostwaldi 1\\
		Tartu, 50411, Estonia\\
		\email{blixt@ut.ee}}

	\author{Mar\'ia-Jos\'e Guzm\'an}
	\address{Departamento de F\'isica y Astronom\'ia, Facultad de Ciencias, Universidad de La Serena,\\ 
	Av. Juan Cisternas 1200, 1720236 La Serena, Chile\\
	Laboratory of Theoretical Physics, Institute of Physics, University of Tartu, W. Ostwaldi 1\\
		Tartu, 50411, Estonia\\
		\email{maria.j.guzman.m@gmail.com}}

			\author{Manuel Hohmann}
	\address{Laboratory of Theoretical Physics, Institute of Physics, University of Tartu, W. Ostwaldi 1\\
		Tartu, 50411, Estonia\\
		\email{manuel.hohmann@ut.ee}}

			\author{Christian Pfeifer}
	\address{Laboratory of Theoretical Physics, Institute of Physics, University of Tartu, W. Ostwaldi 1\\
		Tartu, 50411, Estonia\\
		\email{christian.pfeifer@ut.ee}}

	\maketitle

	\begin{history}
		\received{(Day Month Year)}
		\revised{(Day Month Year)}
	\end{history}

	\begin{abstract}
		We review different approaches to the Hamiltonian analysis of teleparallel theories of gravity. In particular, the Hamiltonian analysis for $f(\mathbb{T})$ theories led to disputed results in the literature. The aim of this review is to relate the different notations and assumptions in the different approaches in a comprehensive way, so that they can be compared more easily. To do this we present the primary constraints of the $f(\tngr)$ gravity class of theories for the first time. The particular cases studied in the literature, $f(\mathbb{T})$ gravity and new general relativity, are contained in this parent theory. We compare their Hamiltonian  analyses done by different authors in the literature among each other by relating them to our analysis of $f(\tngr)$ in detail.
	\end{abstract}


\section{Introduction}

In the recent years, the geometric foundations of general relativity (GR) have been reassessed, and it has been highlighted that its commonly known formulation in terms of the curvature of spacetime is not unique. Equivalently, GR can be formulated in terms of a flat, metric compatible connection with torsion, called the teleparallel equivalent to general relativity (TEGR) or in terms of a flat, torsion-free connection that is not metric compatible, called the symmetric teleparallel equivalent to general relativity (STEGR) \cite{BeltranJimenez:2019tjy}.

From there on, numerous modified theories of gravity have been constructed to overcome the shortcomings of GR such as not explaining the dark matter and dark energy phenomenology, not being consistently quantizable and predicting singularities \cite{Bertone:2004pz,Riess:1998cb,Perlmutter:1998np,Peebles:2002gy,Copeland:2006wr,Weinberg1989,Kiefer:2005uk,Corbelli:1999af,Clowe:2006eq}.

To understand the properties of the theories of gravity beyond GR based on its teleparallel or symmetric teleparallel formulation, it is crucial to have a proper understanding of their canonical structure. Applying the Hamiltonian formalism to these theories allows a nonperturbative counting of the physical degrees of freedom, it states the well-posedness of the Cauchy problem, and can shed some light on canonical quantization.

In this paper, we investigate and review the canonical structure of the most famous teleparallel generalizations of general relativity, so-called $f(\mathbb{T})$ theories \cite{Ferraro:2006jd} and new general relativity (NGR) \cite{Hayashi:1979qx}. Both kinds can be studied collectively by considering $f(\tngr)$ teleparallel theories of gravity.

In the literature, there are several approaches to the Hamiltonian analysis of TEGR \cite{Maluf:1994ji,Maluf:1998ae,Maluf:2000ah,Blagojevic:2000qs,Blagojevic:2000pi,Blagojevic:2000xd,Maluf:2001rg,daRochaNeto:2011ir,Okolow:2013lwa, Maluf:2013gaa,Ferraro:2016wht,doi:10.1142/S0217751X89000704}, of $f(\mathbb{T})$ gravity \cite{Li:2011rn,Ferraro:2018tpu,Blagojevic:2020dyq}\ and NGR and special cases of it  \cite{Cheng:1988zg,Okolow:2011np,Blixt:2018znp,Hohmann:2019sys,Blixt:2019mkt,Mitric:2019rop,Guzman:2020kgh}, in which slightly different definitions of the Lagrangians, the canonical momenta as well as the primary and secondary constraints, and many different notations for all appearing quantities are in use. The approaches differ in how they perform the canonical analysis of the theories, i.e.\ employing an ADM decomposition or not (whose necessity one can already discuss on the level of GR \cite{Kiriushcheva:2008sf}), using tensor components or differential forms and how the gauge freedom encoded in the spin connection is taken into account.

In particular, for $f(\mathbb{T})$ gravity, these studies come to different results on the number of physical degrees of freedom. It is believed that extra degrees of freedom should appear from the breaking of local Lorentz invariance, although the machinery on how this works is complex and requires special care to be taken in its application. There is evidence that some partial or total violation of Lorentz invariance occurs for some circumstances that remain to be studied. The constraint algebra of $f(\mathbb{T})$ gravity is very involved and the matrix of Poisson brackets among constraints presents a variable rank. As a consequence, the number of d.o.f. might not be uniquely defined independent of the field configuration, therefore the disagreement about its number. More details on the current status of the discussion will be presented in Secs. \ref{ssec:hessian} and \ref{ssec:pbmatrix}. The evidence points most probably to 5 d.o.f. in the most general case \cite{Li:2011rn,Blagojevic:2020dyq}, for Minkowski and Friedmann-Lemaitre-Robertson-Walker (FLRW) spacetimes, there is contradictory claims whether it should be 3 d.o.f.  \cite{Ferraro:2018tpu} or 2 d.o.f. as in GR \cite{Blagojevic:2020dyq}, and some far-fetched cases could give 4 d.o.f. or even zero \cite{Blagojevic:2020dyq}. Despite all controversy, $f(\mathbb{T})$ gravity seems an intriguing toy model to build, hopefully, more healthy modified teleparallel gravities.

Less work has been committed to new general relativity. The name of this theory was introduced in \cite{Hayashi:1979qx} as a one-parameter theory agreeing with solar system tests of gravity. However, in this review, we refer to new general relativity as the most general parity even teleparallel gravity theory quadratic in the torsion components. In particular, the one-parameter theory is the most general of the NGR-theories different from TEGR while avoiding pathologies of mixing symmetric and antisymmetric perturbations which is nicely shown in \cite{Jimenez:2019tkx} (see \cite{Kuhfuss1986} for earlier works, and \cite{VanNieuwenhuizen:1973fi} regarding the pathology in general). In \cite{Cheng:1988zg} it is found that this one-parameter theory has a non-deterministic evolution for certain initial values. In \cite{Mitric:2019rop} the Hamiltonian analysis of the NGR-theory with minimal amount of primary constraints was carried out and a special case of this theory was studied in \cite{Okolow:2011nq}. They found that the constraint algebra close without any introduction of secondary constraints (except for the Hamiltonian and momenta constraints generic for any teleparallel theory of gravity). Furthermore, the Hamiltonian analysis for general NGR has been partly carried out in \cite{Blixt:2018znp,Guzman:2020kgh,Hohmann:2019sys,Blagojevic:2020dyq}.
The disagreement in the conclusions which are drawn from the canonical analysis of modified teleparallel theories of gravity by different authors motivates us to presents a comparison of the approaches, and how they can be translated into each other. To do this, we present the primary constraints of the parent class of theories, $f(\tngr)$ gravity, which includes $f(\mathbb{T})$ gravity and NGR as special cases, as reference. We then compare the existing approaches in the literature against our findings and discuss how they are related among each other. We hope this work will simplify the comparison between different approaches to the canonical analysis of modified teleparallel theories of gravity, and thus enable the community to come to a definite answer on the number and nature of the degrees of freedom in these theories.

This work is organized as follows. We perform an introduction to the Dirac-Bergmann algorithm in Sec. \ref{sec:Hamiltonian}. In Sec. \ref{sec:Covariant} we introduce the covariant formulation of the teleparallel formalism and TEGR. Some important points to consider in the Hamiltonian analysis are discussed in Sec. \ref{sec:Points}. In Sec. \ref{sec:fNGR} we introduce $f(\tngr)$ gravity, its primary constraints and their classification. We make a compilation on the different notations and primary constraints for $f(\mathbb T)$ gravity and NGR that can be found in the literature in Sec. \ref{sec:Dictionary}. A discussion on the difficulties in applying the Dirac-Bergmann algorithm can be found in Sec. \ref{sec:Discussion}. Finally, our outlook and conclusions are in Sec. \ref{sec:Conclusions}.

%--------------------------------------Dirac-Bergmann-----------------------------------

\section{Dirac-Bergmann algorithm for Hamiltonian analysis}
\label{sec:Hamiltonian}

Most theories of physical interest are gauge systems: the equations of motion do not determine all the dynamical variables, since there are relations among them that leave the state of the system unaltered. Such situation translates as a constrained Hamiltonian system; in this picture, the canonical variables are not all independent. All gauge systems can be regarded as constrained Hamiltonian systems, but not all constraints from a Hamiltonian system arise from a gauge invariance. Hamiltonian systems with constraints can be studied through the Dirac-Bergmann\footnote{Also called Rosenfeld-Dirac-Bergmann algorithm, see \cite{Rosenfeld:2017umg} and the discussion on Rosenfeld's contribution in \cite{Kiriushcheva:2008sf}} algorithm. In what follows, we will review this method \cite{dirac1964,Sundermeyer:1982gv,Henneaux:1992ig} and put emphasis on some of its peculiarities \cite{Date:2010xr,Sundermeyer:2014kha,Lusanna:2017qsu}. The main steps of the Dirac-Bergmann algorithm are highlighted throughout the text in concordance with the notation introduced in Fig. \ref{HamiltonianTikz}.

In the following, we will introduce the Lagrangian formulation of a theory with finite degrees of freedom, that is a finite number of coordinates depending on time $q^i=q^i(t)$, which define the state of the system. However, later we will study gravitational theories, on which the fields depend on the space-time coordinates. Each field represents, strictly speaking, infinite degrees of freedom. The counting convention therefore is that each field component is called a single physical degree of freedom.

We give a brief introduction to the Lagrange formalism in Sec.~\ref{ssec:lagform}. The canonical momenta and primary constraints are defined in Sec.~\ref{ssec:canmomenta}. The Dirac-Bergmann algorithm and determining the constraint surface is laid out in Sec.~\ref{ssec:diracalgo}. Finally, we define the notions of first and second class constraints in Sec.~\ref{ssec:constraints}. The algorithm is summarized in Fig.~\ref{HamiltonianTikz}.

\tikzstyle{decision} = [diamond, draw, fill=blue!20,
text width=4.5em, text badly centered, node distance=2cm, inner sep=0pt]
\tikzstyle{block} = [rectangle, draw,
text width=7em, text centered, rounded corners, minimum height=3em]
\tikzstyle{ma} = [rectangle, draw,
text width=6em, text centered, rounded corners, minimum height=2.5em]
\tikzstyle{bg} = [rectangle, draw,
text width=6em, text centered, rounded corners, minimum height=2.5em]
\tikzstyle{tbg} = [rectangle, draw,
text width=7em, text centered, rounded corners, minimum height=2.5em]
\tikzstyle{novel} = [rectangle, draw,
text width=6em, text centered, rounded corners, minimum height=2.5em]
\tikzstyle{ngr} = [rectangle, draw,
text width=7em, text centered, rounded corners, minimum height=2.5em]
\tikzstyle{primH} = [rectangle, draw,
text width=8em, text centered, rounded corners, minimum height=2.5em]
\tikzstyle{phidot} = [rectangle, draw,
text width=9em, text centered, rounded corners, minimum height=2.5em]
\tikzstyle{mink} = [rectangle, draw, fill=blue!32,
text width=6em, text centered, rounded corners, minimum height=2.5em, node distance=15cm]
\tikzstyle{line} = [draw, -latex']

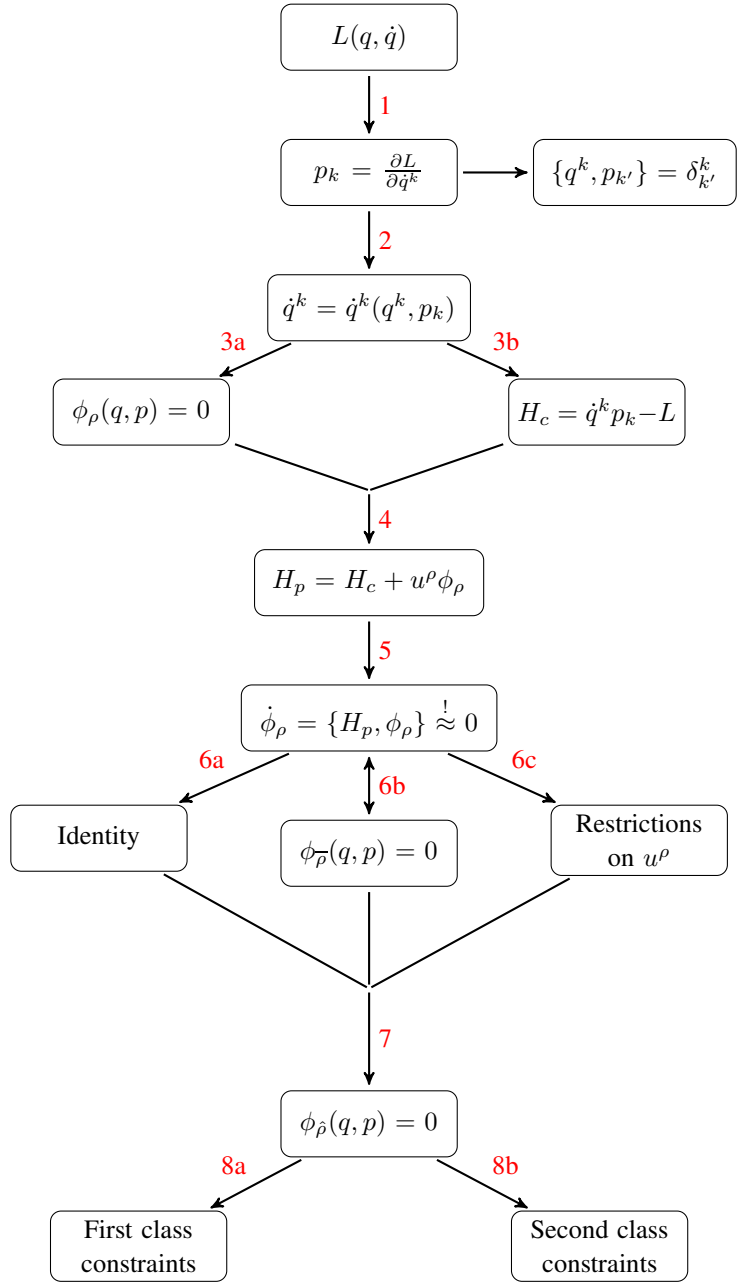
\begin{figure}[htbp!]
\begin{center}
\begin{tikzpicture}[node distance = 1.8cm, auto]
	\node [ma] (L) { $L(q,\dot{q})$ };
	\node [bg, below of =L] (TEGR) {$p_k=\frac{\partial L}{\partial \dot{q}^k}$};
	\node [tbg, right=1cm of TEGR] (Poisson) {$\{q^k,p_{k'}\}=\delta^k_{k'}$};
	\node [ngr, below of =TEGR] (STEGR) {$\dot{q}^k=\dot{q}^k(q^k,p_k)$};
	\node [novel, below left=0.7cm of STEGR] (Prim) {$\phi_{\rho}(q,p)=0$};
	\node [novel, below right=0.7cm of STEGR] (H) {$H_c=\dot{q}^k p_k-L$};
	\node[below=2cm of STEGR, outer sep=-1pt, inner sep=-1pt](4node){};
	\node [primH, below=0.8cm of 4node] (Hp) {$H_p=H_c+u^{\rho}\phi_{\rho}$};
	\node [phidot, below of =Hp] (Vanish) {$\dot{\phi}_{\rho}=\{H_p,\phi_{\rho}\}\overset{!}{\approx}0$};
	\node [novel, below left=1cm of Vanish] (ID) {Identity};
	\node [novel, below of =Vanish] (SC) {$\phi_{\overline{\rho}}(q,p)=0$};
	\node [novel, below right=1cm of Vanish] (Restr) {Restrictions on $u^{\rho}$};
	\node[below of =SC, outer sep=-1pt, inner sep=-1pt] (7node){};
	\node [novel, below of =7node] (Complete) {$\phi_{\hat{\rho}}(q,p)=0$};
	\node [novel, below left=1cm of Complete] (First class) {First class constraints};
	\node [novel, below right=1cm of Complete] (Second class) {Second class constraints};
	\draw[pil,->] (L) -- node  {\hyperlink{step1}{1}} (TEGR);
	\draw[pil,->] (TEGR) -- node  {\hyperlink{step2}{2}} (STEGR);
	\draw[pil,->] (TEGR) -- node {{}} (Poisson);
	\draw[pil,<-] (Prim) -- node {\hyperlink{step3a}{3a}} (STEGR); %Make longer
	\draw[pil,->] (STEGR) -- node {\hyperlink{step3b}{3b}} (H); %Make longer
	\draw[pil,-] (H) -- node {{}} (4node);
	\draw[pil,-] (Prim) -- node {{}} (4node);
	\draw[pil,->] (4node) -- node {\hyperlink{step4}{4}} (Hp); %Make shorter
	\draw[pil,->] (Hp) -- node {\hyperlink{step5}{5}} (Vanish);
	\draw[pil,<-] (ID) -- node {\hyperlink{step6a}{6a}} (Vanish);
	\draw[pil,->] (Vanish) -- node {\hyperlink{step6c}{6c}} (Restr);
%	\draw[<-] (Vanish) to [out=290,in=85,looseness=1] (SC);
%	\draw[->] (Vanish) to [out=250,in=95,looseness=1] (SC);
%	\draw[<->] (Vanish) -- node {{6b}} (SC);
\draw[pil,<->, bend right=45] (Vanish) -- node {\hyperlink{step6b}{6b}} (SC); %This arrow is more clear, but I failed to bend it
	\draw[pil,-] (ID) -- node {{}} (7node);
	\draw[pil,-] (Restr) -- node {{}} (7node);
	\draw[pil,-] (SC) -- node {{}} (7node);
	\draw[pil,->] (7node) -- node {\hyperlink{step7}{7}} (Complete);
	\draw[pil,<-] (First class) -- node {\hyperlink{step8a}{8a}} (Complete);
	\draw[pil,->] (Complete) -- node {\hyperlink{step8b}{8b}} (Second class);
%	\draw[->] (First class) -- node {{}} (Gauge);
	\end{tikzpicture}
\end{center}
\caption{Dirac-Bergmann algorithm} 	\label{HamiltonianTikz}
\end{figure}

\subsection{Lagrangian formalism}
\label{ssec:lagform}

Let us consider a finite-dimensional system with an $n$-dimensional configuration space $Q$ spanned by the coordinates $q^{i}, i=1, \ldots, n$. The system is described by a Lagrangian $L(q(t),\dot{q}(t))$ without explicit dependence on time $t$, which defines the action $S=\int_{t_i}^{t_f} dt L(q,\dot{q})$. The variational principle, i.e.\ the requirement of the action to be stationary
\begin{equation}
    \delta S = \int_{t_i}^{t_f} dt \dfrac{\delta L}{\delta q^{i}(t)} \delta q^{i}(t)=0
\end{equation}
under variations vanishing at the endpoints $t_i, t_f$, yields that physical trajectories $q^{i}(t)$ must satisfy the Euler-Lagrange equations
\begin{equation}
\label{ELeq}
\dfrac{\delta L}{\delta q^{i}} = \dfrac{\partial L}{\partial q^{i} } - \dfrac{\dd}{\dd t}\left( \dfrac{\partial L}{\partial \dot{q}^{i} } \right) = 0.
\end{equation}
Expanding the time derivative in \eqref{ELeq}, we obtain
\begin{eqnarray}
\left( \dfrac{\partial^2 L}{\partial \dot{q}^k \partial \dot{q}^j } \right) \ddot{q}^{j} + \left( \dfrac{\partial^2 L}{\partial \dot{q}^k \partial q^j } \right) \dot{q}^{j} - \dfrac{\partial L}{\partial q^{k} } & = & 0, \\
 V_k - W_{kj} \ddot{q}^{j} & = & 0,
\end{eqnarray}
where $V_k = \frac{\partial L}{\partial q^{k}} - \left( \frac{\partial^{2}L}{\dot{q}^{k} \dot{q}^{j} } \right)\dot{q}^j$ and $W_{kj}=\left( \dfrac{\partial^2 L}{\partial \dot{q}^k \partial \dot{q}^j } \right)$. The object $W_{kj}$ is the Hessian of $L$ with respect to the velocities $\dot q^k$, and it has an important role. If the rank of the Hessian $\text{Rank}(W_{ij})=r=n$, then all accelerations $\ddot{q}^{i}$ can be solved in terms of $q^{i}$ and $\dot{q}^{i}$.
For a constrained physical system, the Hessian has non-maximal rank, that is $\text{Rank}(W_{ij})=r<n$. This means that not all accelerations $\ddot{q}^{i}$ can be uniquely determined in terms of $q^{i}$ and $\dot{q}^{i}$. In the Hamiltonian picture, this implies the existence of primary constraints.

We now recall the steps of the Dirac-Bergmann algorithm needed to obtain the constrained Hamiltonian formulation of the dynamics of the physical system described by the Lagrangian $L$.

%
%
%
%
%
%-------------------------------------------------Canonical momenta and PC-------------------------------------------------------
%
%
%
%
%

\subsection{Canonical momenta and primary constraints}
\label{ssec:canmomenta}

\hypertarget{step1}{\textbf{Step 1}}. For going to the Hamiltonian formalism of a dynamical system, we start with the definition of the canonical momenta as functions of $(q,\dot q)$
\begin{equation}
 p_k(q,\dot{q}) = \dfrac{\partial L}{\partial \dot{q}^k} = \tilde \phi_k(q,\dot q), \qquad k = 1, \ldots, n.
  \label{pcan}
\end{equation}

The fact that the Hessian $W_{ij}=\frac{\partial p_i}{\partial \dot{q}^j}$ has non-maximal rank $r$ implies that it has a non-trivial kernel of dimension $n-r$. This kernel is spanned by $n-r$ vectors with components $\kernelcomp^k_{\rho}$ such that
\begin{equation}
 \kernelcomp^k_{\ \rho} W_{kj} = 0, \qquad  \rho = 1, \ldots, n - r.
\end{equation}

\hypertarget{step2}{\textbf{Step 2}}.
When the rank of $W_{kj}$ is $r<n$ (also that $\text{det}(W_{kj})=0$), we can solve $r$ velocities in terms of the momenta and positions from Eq.\eqref{pcan},
\begin{equation}\label{solvveloc}
\dot{q}^{ \hat{i}} = {\mathcal{F}}^{\hat{i}}(q,p_{\alpha} ,\dot{q}^{\overline{i} } ),
\end{equation}
where, without loss of generality, it can be assumed that $\hat{i} = 1, \ldots, r,$ label the solvable velocities, and $\overline{i} = r+1, \ldots, n$ are the velocities that can not be solved. We also assume that the index $\alpha$ can take $r$ values. If we substitute \eqref{solvveloc} in \eqref{pcan}, we obtain
\begin{equation}\label{eq:prim}
p_k = \tilde{ \phi }_k(q, \dot{q}^{\hat{i}} , \dot{q}^{\overline{i}}) = \tilde{ \phi }_k(q, \mathcal{F}^{\hat{i} }(q, p_{\alpha}, \dot{q}^{\overline{i} } ),\dot{q}^{\overline{i}} ) = \hat \phi_k(q, p_{\alpha}, \dot{q}^{\overline{i}} )
\end{equation}
The functions $\hat \phi_k(q, p_{\alpha}, \dot{q}^{\overline{i}})$ cannot depend on $\dot{q}^{\overline{i} }$ any longer, otherwise it would be possible to solve for more of the velocities.

\hypertarget{step3a}{\textbf{Step 3a}}.
For $k=1,\ldots,r$ the equations $p_k = \hat \phi_k$, \eqref{eq:prim}, are trivially satisfied, from the definition of the momenta and the assumption of being able to express the velocities $\dot q^{\hat i}$ as functions of $q$ and $p$, while for $k=r,\ldots,n$ one obtains $n-r$ non-trivial relations $p_{\rho}=\hat \phi_{\rho}(q,p_{\alpha})$ that relate coordinates and momenta. These give rise to the so-called primary constraints
\begin{equation}
 \phi_{\rho}(q,p)  = p_\rho - {\hat \phi}_\rho= 0, \ \ \ \rho = 1, \ldots, n-r.
\label{vprim}
\end{equation}
An important notion for the Hamiltonian formalism is ``weak equality'' (denoted by the symbol ``$\approx$'' ) which is an equality on the constraint surface. The symbol ``$\accentset{!}{\approx}$'' will denote that weak equality is imposed, which means that the equality of both sides of the equations is not necessarily implied by the previously found constraints, but must be imposed as an additional condition in order to obtain the final constraint surface.

\hypertarget{step3b}{\textbf{Step 3b}}.
The canonical Hamiltonian can be obtained in terms of the momenta as
\begin{equation}
    H_c = \dot{q}^{i} p_{i} - L(q,\dot{q}),
\end{equation}
which can be proved to depend only on $q^{i}$ and $p_{i}$, not in the velocities $\dot{q}^{i}$. This Hamiltonian does not encode \textit{a priori} the primary constraints and thus does not describe the same dynamical system as the Lagrangian. To keep the predictions in the transition from the Lagrangian to the Hamiltonian formulation of the dynamics of the system unaltered, one needs to add the primary constraints with the help of Lagrange multipliers $H_c \longrightarrow H_c + u^{\rho} \phi_{\rho}(q,p)$.

\hypertarget{step4}{\textbf{Step 4.}} We consider the primary Hamiltonian
\begin{equation}
    H_p(q,p) = H_c + u^{\rho} \phi_{\rho}(q,p),
\end{equation}
where the $u^{\rho}$ are Lagrange multipliers (arbitrary functions); they ensure the primary constraints from the beginning. It can be shown that $H_p$ generates the time evolution of the physical system through the Poisson brackets (PB) in the following way. For any function $F$ in the phase space, it is
\begin{equation}
\label{timev}
\dot{F}=\{F,H_p \} \approx \{F,H_c \} + u^{\rho}\{F,\phi_{\rho} \},
\end{equation}
where the term $\{F, u^{\rho} \} \phi_{\rho}$ is dropped since it vanishes weakly, and the PB between two functions $F(p,q)$ and $G(p,q)$ are defined as
\begin{equation}
\{F,G \} = \dfrac{\partial F}{\partial q^{i}} \dfrac{\partial G}{\partial p_i } - \dfrac{\partial F}{\partial p_i} \dfrac{\partial G}{\partial q^{i}}.
\end{equation}
This definition allows to compute the time evolution of any primary constraint \eqref{timev}, whose outcome has many different branches, as we explain in the next section.

%
%
%
%
%---------------------------------------------------------------Dirac-Bergmann----------------------------------------------------
%
%
%
%
%

\subsection{Dirac-Bergmann algorithm and determination of constraint surface}
\label{ssec:diracalgo}

\hypertarget{step5}{\textbf{Step 5.}} In order for the physical system to be consistent, the evolution on time $\dot{\phi}_{\rho}$ of primary constraints should be zero. This imposes the condition
\begin{equation}
\label{ccond}
\dot{\phi}_{\rho} = \{ \phi_{\rho}, H \} + u^{\sigma}\{\phi_{\rho}, \phi_{\sigma} \} \overset{!}{\approx} 0.
\end{equation}
If we define $h_{\rho}= \{ \phi_{\rho},H \}$ and $C_{\rho\sigma}=\{ \phi_{\rho}, \phi_{\sigma} \}$, then \eqref{ccond} can have two outcomes, depending if $\text{det}(C)$ is weakly zero or not.\\

\hypertarget{step6a}{\textbf{Step 6a.}} If $\text{det}(C) \approx 0$, then the multipliers are not uniquely determined, and \eqref{ccond} is only solvable if the $h_{\rho}$ satisfies the conditions
    \begin{equation}
        \omega^{\rho}_{\alpha} h_{\rho} \overset{!}{\approx} 0,
    \end{equation}
	where $\omega^{\rho}_{\alpha}$ are $p-m$ linearly independent vectors spanning the kernel of $C$, which has rank $m$.\\

\hypertarget{step6b}{\textbf{Step 6b.}} These conditions can be fulfilled like in the previous step, or lead to a certain number $s'$ of new constraints
    \begin{equation}
        \phi_{\overline{\rho}} \approx 0, \ \ \ \overline{\rho} = n-r+1, \ldots, n-r+s'
    \end{equation}
    called secondary constraints.\\

\hypertarget{step6c}{\textbf{Step 6c.}} If $\text{det}(C) \not \approx 0$, Eq.\eqref{ccond} is an inhomogeneous system of linear equations with solutions
    \begin{equation}
        u^{\sigma} \approx -(C^{-1})^{\rho\sigma} h_{\rho},
    \end{equation}
    and the Hamilton equations of motion for a function $F(p,q)$ become
    \begin{equation}
        \dot{F} \approx \{ F, H_c \} - \{ F, \phi_{\rho} \} (C^{-1})^{\rho\sigma} \{ \phi_{\sigma}, H_c \},
    \end{equation}
    which do not contain arbitrary multipliers, that is, they are fully determined.

Unlike primary constraints, secondary constraints have been derived from the equations of motion. The procedure of vanishing the time evolution should be iterated with secondary constraints, which could give rise to tertiary constraints and so on, until no more constraints appear. In most cases of physical relevance, the algorithm terminates at the stage of secondary constraints, but it is not hard to build pathological examples on which there is an infinite tower of constraints, or the conditions of time consistency give rise to physically inequivalent branches \cite{Henneaux:1992ig}. We could also face the unlikely case on which the consistency conditions are incompatible with each other, then it is said that the Hamiltonian system is inconsistent, and the algorithm is terminated \cite{Date:2010xr}

\hypertarget{step7}{\textbf{Step 7.}} When no more constraints appear, we are left with a hypersurface defined by
\begin{eqnarray}
\phi_{\rho} \approx 0, & & (\rho=1,\ldots,n-r),\\
\phi_{\overline{\rho}} \approx 0, & & (\overline{\rho}=n-r+1,\ldots,n-r+s).
\end{eqnarray}
The first set $\{\phi_{\rho} \}$ contains all $p$ primary constraints, while the set $\{\phi_{\overline{\rho} } \}$ contains $s$ secondary, tertiary, etc. constraints. By using a common notation for all constraints as $\phi_{\hat{\rho}}$, with $\hat{\rho}=1,\ldots,n-r+s$, we can define the matrix of constraints as
\begin{equation}
C_{\hat{\rho}\rho } = \{\phi_{\hat{\rho} }, \phi_{\rho}  \}.
\end{equation}
If $\omega^{\hat{\rho}}_{\alpha}$ span the left kernel of $C_{\hat{\rho}\rho }$, then the conditions $\omega^{\hat{\rho} }_{\alpha} \{\phi_{\hat{\rho}},H_c \} \approx 0$ are satisfied. Also for the multipliers, the equations
\begin{equation}
\label{multeq}
\{ \phi_{\overline{\rho}}, H_c \} + \{ \phi_{\overline{\rho}}, \phi_{\rho} \} u^{\rho} \approx 0
\end{equation}
are fulfilled. Note that the weak equalities are defined with respect to the final constraint hypersurface of all  constraints.

%
%
%
%
%------------------First and Secondary constraints-------------------------------------------
%
%
%
%
%

\subsection{First and second class constraints}
\label{ssec:constraints}

Solving the multiplier functions from \eqref{multeq} leads to the definition of first and second class constraints. If the rank of the matrix $C_{\hat{\rho}\rho }$  is $n-r$, then all multipliers are fixed, but if its rank is $k<n-r$, there are $n-r-k$ linearly independent solutions of the equation
\begin{equation}
C_{\hat{\rho}\rho} V^{\rho}_{\alpha} = \{ \phi_{\hat{\rho} }, \phi_{\rho}  \} V^{\rho}_{\alpha} \approx 0,
\end{equation}
which is the homogeneous part of \eqref{multeq}. Notice that $V^{\rho}_{\alpha}$ span the right kernel of $C_{\hat{\rho}\rho}$. With all this, the most general solution of \eqref{multeq} is a sum of a particular solution $U^{\rho}$ and a linear combination of the solutions of the homogeneous part, that is
\begin{equation}
u^{\rho} = U^{\rho} + v^{\alpha} V^{\rho}_{\alpha},
\end{equation}
where the coefficients $v^{\alpha}$ are arbitrary.

It is important to keep in mind that the rank of $C_{\hat{\rho}\rho}$ can be variable, and in such case, it can still give rise to different, but consistent physical evolution. This is not always the case, and some counterexamples can be found at \cite{Henneaux:1992ig}. This feature seems to be crucial for modified teleparallel gravities.

\hypertarget{step8a}{\textbf{Step 8a.}} We define a function $\mathcal{F}(p,q)$ in the phase-space to be first class if the PB with all constraints in the theory vanish,
\begin{equation}
\{ \mathcal{F}(p,q), \phi_{\hat{\rho} } \} \approx 0.
\end{equation}

\hypertarget{step8b}{\textbf{Step 8b.}} If a phase-space function is not first class, it is called to be second class.

Since the PB satisfy the Jacobi identity, it is possible to prove that the PB of two first class constraints are itself first class. It is convenient to reformulate a theory in terms of its maximal number of independent first class and second class constraints. Let us assume that the maximal number of first class constraints is obtained after building some linear combination, which we will denote as $\Phi_{I}, I=1, \ldots, l$, and the remaining set of second class constraints is $\chi_A$. Then, to make sure that the maximum number of $\Phi_{I}$ has been found, it is convenient to build the PB matrix of second class constraints
\begin{equation}
\Delta_{AB} = \{ \chi_A , \chi_B \}
\end{equation}
and check that it has non-vanishing determinant. After this, we make sure that $\Delta_{AB}$ has an inverse, and that the Lagrange multipliers for second class constraints can be solved univocally. After all this procedure, it is possible to count the physical degrees of freedom of the theory through the formula
\begin{equation}
\text{Number of d.o.f.} = \text{Number of}\ (p,q) - \text{Number of f.c.c.} - \dfrac12 (\text{Number of s.c.c.}).
\end{equation}

In summary, the procedure of the Dirac-Bergmann algorithm has two main goals: first, finding all primary, secondary, tertiary, etc., constraints from the definition of the canonical momenta and the time evolution of the system, and second, grouping them into first and second class constraints.

%
%
%
%
%--------------------------------Covariant formulation-----------------------------
%
%
%
%
%

\section{Covariant formulation of teleparallel gravity}
\label{sec:Covariant}

In the following, a covariant formulation for teleparallel gravity will be introduced and notation and conventions for this paper will be fixed. Greek letters $\mu,\nu,\rho,\ldots$ denote spacetime indices, Lorentz tangent space indices are denoted by the first letters of the Latin alphabet $A,B,C,\ldots$, and their spatial part is denoted with hats $\hat{A},\hat{B},\hat{C},\ldots$. The sign convention for the Minkowski metric is the mostly negative one: $\eta_{AB}=\mathrm{diag}(1,-1,-1,-1)$. The torsion components are defined as
\begin{align}
    T^A{}_{\mu\nu}=\partial_\mu \tetrad^A{}_\nu-\partial_\nu \tetrad^A{}_\mu +\spinconnection^A{}_{B\mu }\tetrad^B{}_\nu-\spinconnection^A{}_{B\nu }\tetrad^B{}_\mu,
\end{align}
where $\tetrad^A{}_\mu$ are tetrad components and $\spinconnection^A{}_{B\mu }$ are components of the spin connection defined as
\begin{align}\label{eq:spinconnlor}
    \spinconnection^A{}_{B\mu} = -\left(\Lorentz^{-1}\right)^C{}_B \partial_\mu \Lorentz_C{}^A,
\end{align}
where $\Lorentz_C{}^A$ are Lorentz matrices. Taking into account this spin connection and applying Lorentz transformations simultaneously to spin connection and tetrad only, yields that the torsion tensor transform covariantly and the formulation of teleparallel gravity, in this sense, satisfy local Lorentz invariance \cite{Krssak:2015oua,Krssak:2018ywd}.

Cotetrads are denoted by $\cotetrad_A{}^\mu$ and the following relations are satisfied
\begin{align}
    \eta_{AB}=\metric^{\mu\nu}\cotetrad_A{}^\mu \cotetrad_B{}^\nu,
\end{align}
\begin{align}
    \metric_{\mu\nu}=\eta_{AB}\tetrad^A{}_\mu \tetrad^B{}_\nu.
\end{align}
Lorentz indices and spacetime indices can be transformed between each other by contraction with a tetrad resp. cotetrad in the obvious correct way, i.e\ a spacetime index ${}^\mu$ becomes a Lorentz index ${}^A$ through contraction with a tetrad $\theta^A{}_\mu$, while a spacetime index ${}_\mu$ becomes a Lorentz index ${}_A$ through contraction with an inverse tetrad $e_A{}^\mu$. Lorentz indices are raised and lowered with the Minkowski metric, while spacetime indices are raised and lowered with the spacetime metric.

The teleparallel equivalent to general relativity (TEGR) is obtained from rewriting the classical Einstein-Hilbert action of general relativity in the teleparallel geometric language, and yields the action
\begin{align}
\label{eq:Action}
S = \int \dd^4x L_{\mathrm{TEGR}} + S_{\text{M}}\,,
\end{align}
where the Lagrangian is given by
\begin{align}
    L_{\mathrm{TEGR}}
    &=\frac{1}{2 \kappa} \dettetrad \mathbb{T} = \frac{1}{4 \kappa} \dettetrad  T^{\rho\mu\nu}S_{\rho\mu\nu} \\
    &=\frac{1}{2 \kappa} \dettetrad \left(\frac{1}{4}T^\rho{}_{\mu\nu}T_\rho{}^{\mu\nu}-\frac{1}{2}T^\rho{}_{\mu\nu}T^{\mu\nu}{}_\rho-T^\rho{}_{\mu\rho}T^{\sigma\mu}{}_\sigma \right),
\end{align}
$\kappa=\frac{8\pi G}{c^4}$, $\dettetrad$ is the determinant of the tetrad and $S_{\rho\mu\nu}$ is the so called superpotential
\begin{align}\label{eq:sp}
    S_{\rho\mu\nu} = \frac{1}{2} T_{\rho\mu\nu} + T_{[\nu\mu]\rho} + 2 g_{\rho[\mu}T^\sigma{}_{\nu]\sigma}\,.
\end{align}
The scalar $\mathbb{T}$ is called the canonical torsion scalar. It is related to the Ricci scalar $\lc{R}$ of the Levi-Civita connection of the metric generated by the tetrad by a total derivative term
\begin{align}
\mathbb{T} = \frac{1}{\sqrt{\det g}}\partial_\mu ( \sqrt{\det g}T_\nu{}^{\nu\mu}) - \lc{R}\,,
\end{align}
which is used to prove the dynamical equivalence between TEGR and GR \cite{Maluf:2013gaa,AP}.
%
%
%
%
%----------------------------Things to keep in mind----------------------------------
%
%
%
%
%

\section{Aspects of Hamiltonian analysis of teleparallel gravity}
\label{sec:Points}
We now state relevant, preliminary steps which are necessary for the Hamiltonian analysis of teleparallel gravity theories. First, we discuss different possible choices for the fundamental fields and gauge choices in Sec.~\ref{ssec:fields}, then in Sec.~\ref{ssec:31decomp} we discuss how to establish the proper split of the teleparallel geometry into space and time components.
%
%
%
%----------------------------------Fundamental fields----------------------------------
%
%
%
%
%

	\subsection{Fundamental fields and gauge fixing}
	\label{ssec:fields}

	Before starting the Dirac-Bergmann algorithm one should consider another step, which could be labeled as ``step 0'' in the diagram in Fig.~\ref{HamiltonianTikz}. This step comprises identifying the fundamental fields \(q^i\), which constitute the dynamical variables of the theory. Here, it is desirable to reduce the number of constrained variables as much as possible, in order to reduce the dimension of the Hessian and the number of Poisson brackets between constraints to be calculated. In the ideal case, one may parametrize the physical phase space using only independent variables, and obtain a system without constraints. Even if this is not always possible, one may usually reduce the number of constrained variables by a change of parametrization. Nevertheless, this reparametrization possibly comes at the cost of a more complicated Lagrangian.

	As an example, one may consider general relativity. In the usual metric formulation, the fundamental fields, which correspond to the generalized coordinates \(q^i\) in Sec.~\ref{sec:Hamiltonian}, are the 10 independent components of the metric. However, it is possible to formulate GR purely in the tetrad formalism. If the tetrad components are used instead of the metric ones, there are six more fundamental fields to be treated in the Hamiltonian analysis. In consequence, there are as well six additional primary constraints corresponding to generators of local Lorentz transformations. They reflect the arbitrariness in the choice of the tetrad for a given metric. Hence, choosing the metric components as fundamental variables reduces the number of constrained variables, compared to the tetrad formulation.

	In teleparallel gravity, the metric components do not suffice as the fundamental fields, due to the formulation of the Lagrangian in terms of the torsion tensor, which cannot be obtained from the metric alone. The most common variables chosen in the covariant formulation of teleparallel gravity displayed in the previous section, which is manifestly invariant under local Lorentz transformations, are the tetrad \(\tetrad\) and the flat Lorentz spin connection \(\omega\). The latter may further be parametrized by finite Lorentz transformations. However, it follows from the Lorentz invariance of the teleparallel gravity action in the covariant formulation that the canonical momenta of the spin connection are related to the momenta of the tetrad, revealing that these spin connection degrees of freedom are not independent~\cite{Blixt:2019mkt}. As a consequence, one may choose a different parametrization, in order to reduce the number of constrained field variables. These different parametrizations give rise to several approaches on how to consider the spin connection as dynamical field in teleparallel gravity,  see Ref.~\cite{Golovnev:2017dox} for an extended discussion. In summary, we have the following choices for our fundamental fields:

	\begin{enumerate}
	\item
	The most straightforward approach is to consider both, the $16$ tetrad components and the $24$ components of a Lorentz spin connection, as fundamental variables. However, since the spin connection components are not independent of each other, but constrained by the flatness condition, in addition a set \(\lambda_{\mu}{}^{\nu\rho\sigma}\) of Lagrange multipliers is required, which enforce the vanishing of the curvature. The full teleparallel action then takes the symbolic form~\cite{Nester:2017wau}
	\begin{equation}
	S_{\text{TG}} = S_{\text{TG}}(\tetrad,\spinconnection) - \int \dd^4x\,\dettetrad\,\lambda_{\mu}{}^{\nu\rho\sigma}\Ricci^{\mu}{}_{\nu\rho\sigma}\,.
	\end{equation}
	\item
	Alternatively, one may implement the flatness of the spin connection by exploiting the relation~\eqref{eq:spinconnlor} in order to express the spin connection in terms of a local Lorentz transformation. In this case, the fundamental variables become the tetrad components and the components of the Lorentz matrices that parametrize the inertial spin connection, so that the action takes the structure $S_{\text{TG}} = S_{\text{TG}}(\tetrad,\Lorentz)$\,.
	\item Finally, one may use the aforementioned approach as a starting point to further reduce the number of fundamental variables, based on the aforementioned observation that the canonical momenta of the spin connection are linearly dependent on the tetrad momenta. This is a consequence of the fact that the spin connection is a pure gauge degree of freedom, so that the Lorentz matrices introduced above enter the action only in the combination \(\tilde{\tetrad}^A{}_{\mu} = \tetrad^B{}_{\mu}\Lorentz_B{}^A\). Hence, one may replace the tetrad \(\tetrad\) in the action by \(\tilde{\tetrad}\), and obtains an action which is independent of \(\Lorentz\); schematically, $S_{\text{TG}} = S_{\text{TG}}(\tetrad,\Lorentz) = S_{\text{TG}}(\tilde{\tetrad})$. This leaves only the $16$ components of the tetrad \(\tilde{\tetrad}\) as fundamental field variables, and is formally equivalent to imposing the Weitzenböck gauge $\spinconnection^A{}_{B\mu} = 0$ (or alternatively $\Lambda_{B}{}^{A} = \text{const}$).
	\end{enumerate}
	Clearly, while all three approaches lead to the same physical phase space, and thus equivalent results, the latter approach introduces the smallest number of fundamental field variables. In the following, we will therefore assume that the Weitzenböck gauge is imposed, leaving the tetrad as only fundamental variable, and not consider the other two approaches. Also, we will drop the tilde in the notation and simply denote the tetrad by \(\tetrad\).

	%
	%
	%
	%
	%----------------------------------------------3+1 decomposition----------------------------
	%
	%
	%
	%
	%

	\subsection{$3+1$ decomposition} 
	\label{ssec:31decomp}

	The Hamiltonian formalism requires a Legendre transformation from the set of fields and their velocities to the fields and their conjugate momenta. To invert the velocity momentum relations, it is very convenient to employ a 3+1 decomposition of spacetime. Geometrically, this mean that we split the $4$-dimensional spacetime into $3$-dimensional hypersurfaces and a time direction. The geometry of such a foliated spacetime can be described in two ways. The first is by introducing adapted coordinates $(x^0,x^i), i=1,2,3$, where $x^i$ denotes intrinsic coordinates on each hypersurface, and the time coordinate $x^0$ labels the hypersurfaces of the foliation, such that each hypersurface is given by setting $x^0=\text{const}$. These coordinates define a coordinate basis \(\partial_{\mu}\) of the tangent space, with \(\partial_i\) being tangent to each hypersurface, and an extrinsic basis vector field \(\partial_0\). Alternatively, instead of the coordinate vector field $\partial_0$ associated with time coordinate, given a metric $g_{\mu\nu}$ one may use the normal vector $\normalvector$ to the hypersurfaces as additional extrinsic reference direction. Both descriptions are related by the so called lapse function $\lapse$ and shift vector $\shift = \shift^i \partial_i$, by expanding the $x^0$ tangent direction as
	\begin{align}
    \label{ADMzero}
    \partial_0=\lapse\normalvector^\mu\partial_\mu +\shift^i\partial_i.
    \end{align}
    From this we can read off that the component of the normal vector to hypersurfaces of constant time slices reads
    \begin{align}
    \normalvector^0=\frac{1}{\lapse}, \indent \normalvector^i=-\frac{\shift^i}{\lapse}.
    \end{align}
	Introducing a negative definite intrinsic metric $\gamma = \gamma_{ij}dx^idx^j$ on the hypersurfaces the metric can be written as
	\begin{align}\label{eq:ADMmetric}
		\metric_{\mu\nu}=\begin{bmatrix}
		\lapse^2+\shift^i\shift^j\inducedmetric_{ij} & \shift_i\\
		\shift_i & \inducedmetric_{ij}
		\end{bmatrix}\,.
	\end{align}
	The variables $\lapse,\shift_i,\inducedmetric_{ij}$ are normally refrerred as ``ADM-variables'', named after R. Arnowitt, S. Deser, and C. W. Misner \cite{PhysRev.116.1322}. It is not difficult to find a tetrad for which the ADM decomposition of the metric is reproduced. For instance,
	 \begin{align}
	 	\tetrad^A{}_\mu=\begin{bmatrix}\lapse & 0  \\
	 	\shift^i\tetrad^{\hat{A}}{}_i & \tetrad^{\hat{A}}{}_i \end{bmatrix},
	 \end{align}
	 yields that $\metric_{\mu\nu}= \tetrad^A{}_\mu\tetrad^B{}_\nu\eta_{AB}$ has the form \eqref{eq:ADMmetric}, where $\inducedmetric_{ij}=\tetrad^{\hat{A}}{}_i\tetrad^{\hat{B}}{}_j\eta_{AB}$.
	 However, a few things must be remarked about this tetrad. First, note that it requires a $3+1$ split into space and time components not only for the spacetime indices, but also for the Lorentz indices. Further, its form is not invariant under (global or local) Lorentz transformations. This implies that the choice of this tetrad explicitly introduces a gauge condition, by imposing certain components of the tetrad to vanish. Fixing the gauge does not go without consequence in teleparallel gravity theories, since in general such theories are not invariant under (local) Lorentz transformations of the tetrad alone. It follows that by imposing gauge conditions on the tetrad, one must allow for a non-vanishing spin connection, in order to restore the Lorentz invariance under simultaneous transformations of the tetrad and the spin connection. The alternative approach, which we favor here, is to impose a gauge condition on the spin connection only, such as the Weitzenböck gauge \(\spinconnection^A{}_{B\mu} \equiv 0\), and to keep all 16 components of the tetrad as unrestricted dynamical variables. Also in this case, an ADM decomposition~\eqref{eq:ADMmetric} can be achieved. The crucial insight is that the spatial metric components \(\inducedmetric_{ij} = \eta_{AB}\tetrad^A{}_i\tetrad^B{}_j\), and in consequence the Lorentz components \(\normalvector^A\) of the unit normal vector, can be fully expressed in terms of the spatial tetrad components \(\tetrad^A{}_i\), and do not involve the time components \(\tetrad^A{}_0\). To express the latter, one realizes that \((\normalvector^A, \tetrad^A{}_i)\) are a linearly independent set of four Lorentz vectors, and hence form a basis of the Minkowski space. One may thus express the Lorentz vector \(\tetrad^A{}_0\) in this basis as
\begin{align}
\tetrad^A{}_0=\lapse\normalvector^A+\shift^i\tetrad^A{}_i\,,
\end{align}
thereby defining the lapse \(\lapse\) and shift \(\shift^i\) as the coefficients of \(\tetrad^A{}_0\) with respect to this basis. One finds that the metric is indeed of the form~\eqref{eq:ADMmetric}, while keeping 16 independent dynamical variables \((\lapse, \shift^i, \tetrad^A{}_i)\) without imposing gauge conditions on them.

%
%
%-----------------------------------Hessian------------------------------------------
%
%
%
%

%
%
%
%
%---------------------------------------------Primary constraints-----------------------------------
%
%
%
%
%

\section{Canonical momenta and primary constraints in $f(\tngr)$}
\label{sec:fNGR}

In the literature of the Hamiltonian analysis of modified teleparallel theories of gravity, the focus lies on two different classes of theories: new general relativity (NGR) \cite{Hayashi:1979qx}, and $f(\mathbb{T})$ gravity \cite{Ferraro:2006jd}. To compare the different approaches in an efficient way later, we newly present here the derivation of the momenta for $f(\tngr)$ teleparallel theories of gravity, from which different primary constraints in different subclasses of the theory arise. $f(\tngr)$ gravity is a special case of the $f(T_\mathrm{ax},T_\mathrm{ten},T_\mathrm{vec})$ gravity theories which were introduced in \cite{Bahamonde:2017wwk}. Phenomenologically this theory has not yet been studied in full detail, but some results about gravitational waves~\cite{Hohmann:2018xnb,Hohmann:2018jso} and its post-Newtonian limit~\cite{Ualikhanova:2019ygl} are known.

The starting point of the analysis is the Lagrange density
\begin{align}\label{Lftngr}
L_{f(\tngr)} = \frac{1}{2\kappa}\dettetrad f(\tngr)\,,
\end{align}
where
\begin{align}
    \tngr = H_{ABC}T^{ABC}\,,
\end{align}
with
\begin{align}\label{eq:induction}
    H_{ABC} = c_1 T_{ABC} + \frac{c_2}{2} (T_{CBA} - T_{BCA}) + \frac{c_3}{2}(\eta_{AB}T^D{}_{DC} - \eta_{AC} T^D{}_{DB})\,.
\end{align}
We call $H_{ABC}$ the NGR induction tensor, in resemblance with induction tensors in electrodynamics \cite{Hohmann:2017duq}, or the NGR superpotential. By fixing the three parameters $c_1, c_2, c_3$ to the specific values
\begin{align}\label{eq:NGRTEGR}
c_1 = \frac{1}{4}\,, \quad c_2=\frac{1}{2}\,, \quad c_3 = -1,
\end{align}
the NGR induction tensor becomes proportional to the usual superpotential \eqref{eq:sp}, as one finds  $H_{ABC} = \frac{1}{2}S_{ABC}\,$, and the Lagrangian \eqref{Lftngr} becomes the standard $f(\mathbb{T})$ gravity Lagrangian.

To perform the Hamiltonian analysis of $f(\tngr)$ gravity, a Legendre transform is performed, in order to obtain a mathematically equivalent formulation of $f(\tngr)$ gravity. This is done with help of a scalar field $\phi(\tngr)$ in the form
\begin{align}
\label{Lfngr}
    L_{\mathrm{f}} =\frac{1}{2\kappa} \dettetrad( \phi \tngr - V(\phi))\,.
\end{align}
The fundamental variables of the theory are now the tetrad $\tetrad^A$ as well as the scalar field $\phi$. This theory reduces to NGR by setting $\phi=1$ and $V(\phi)=0$, and it reduces to $f(\mathbb{T})$ gravity for the TEGR choice of parameters \eqref{eq:NGRTEGR}.

The canonical momenta of these fields are identified as
\begin{align}\label{eq:pifNGR}
      \momenta_A{}^\mu &= \frac{\partial L}{\partial_0 \tetrad^A{}_\mu} = \frac{\partial L}{\partial T^A{}_{0\mu}} = \frac{2}{\kappa} \dettetrad \phi H_A{}^{0\mu}\,, \\
     \frac{\kappa}{2} \pi_\phi &= \frac{\partial L}{\partial \partial_0 \phi} = 0\,.
\end{align}
The momentum equations yield immediately five trivial universal primary constraints
\begin{align}\label{eq:primfngr}
\momenta_A{}^0 = 0\ \textrm{and}\  \pi_\phi = 0\,,
\end{align}
where the first four come from the index symmetries of $H_{ABC}$.

For further analysis, we consider the momenta $\pi_{AB}$ with pure lowered Lorentz indices. These are related to the momenta $\pi_A{}^\mu$ with the canonical index positions as
\begin{align}\label{eq:hatpi}
    \frac{\kappa}{2} \pi_{AB}
    &=  \frac{\kappa}{2} \momenta_A{}^\mu \tetrad^C{}_\mu \eta_{BC} = \phi \dettetrad H_A{}^0{}_B =  \phi \frac{\dettetrad}{\lapse} \normalvector_\mu  H_A{}^\mu{}_B \nonumber \\
    & = \phi \sqrt{\gamma} \normalvector^D H_A{}_{DB} \,,
\end{align}
where we used that in a coordinate basis we have $\normalvector_0 = \lapse$ and $\normalvector_i = 0$.

It satisfies $\normalvector^B \pi_{AB}=0$, which represents the four constraints
\begin{equation}
\label{pcBHP}
\pi_A{}^0 = 0.
\end{equation}
Further, we can decompose the remaining 12 components with respect to the three dimensional rotational group on the equal time hypersurface, also called $\mathcal{VAST}$ decomposition, with help of the projectors $\pr^A_B = \delta^A_B - n_B n^A$ as
\begin{enumerate}
    \item a {$\mathcal{V}$}ectorial part
    \begin{align}\label{eq:vecfngr}
        \frac{\kappa}{2 \phi \sqrt{\gamma}} \normalvector^A \pi_{AB}
        &=  \normalvector^A \normalvector^D H_A{}_{DB}  \nonumber\\
        &= \frac{1}{2}(2 c_1 + c_2) \normalvector^A\normalvector^D T_{ADB} + \frac{c_3}{2} (T^A{}_{AB} - \normalvector_B \normalvector^D T^A{}_{AD})\nonumber\\
        &=\frac{1}{2}(2 c_1 + c_2 + c_3) \normalvector^A\normalvector^D T_{ADB} + \frac{c_3}{2} T^Q{}_{PD} \pr^D_B \pr^P_Q\,;
    \end{align}
    \item an {$\mathcal{A}$}ntisymmetric part
    \begin{align}\label{eq:asymfngr}
        \frac{\kappa}{2 \phi \sqrt{\gamma}}\pr^C_{[A} \pr^D_{B]}\pi_{CD}
        &= \frac{\kappa}{2 \phi \sqrt{\gamma}}(\hat\pi_{[AB]} - \normalvector_{[A} \normalvector^C \hat \pi_{|C|B]})\nonumber\\
        &= \left(\frac{1}{2}(2c_1-c_2) \normalvector^DT_{PDQ} + \frac{c_2}{2}\normalvector^D T_{DPQ}\right)\pr^P_{[A}\pr^Q_{B]}\,;
    \end{align}
    \item a {$\mathcal{S}$}ymmetric trace free part
    \begin{align}\label{eq:symfngr}
        &\frac{\kappa}{2 \phi \sqrt{\gamma}}    (\pr^C_{(A} \pr^D_{B)}\pi_{CD} - \frac{1}{3}(\eta_{AB} - \normalvector_A \normalvector_B)\pi^C{}_C) \nonumber\\
        &= \frac{1}{2}(2c_1+c_2) (\pr^C_{(A} \pr^D_{B)} - \frac{1}{3}(\eta_{AB} - \normalvector_A \normalvector_B) \eta^{CD}) T_{CQD}\normalvector^Q\,;
    \end{align}
        \item a {$\mathcal{T}$}race part
    \begin{align}\label{eq:trfngr}
        \frac{\kappa}{2 \phi \sqrt{\gamma}} \eta^{AB}\pr^C_{A} \pr^D_{B}\pi_{CD}
        &= \frac{\kappa}{2 \phi \sqrt{\gamma}} \pi^A{}_A \nonumber\\
        &= \frac{1}{2}(2c_1 + c_2 + 3 c_3)\normalvector^D T^A{}_{DA}\,.
    \end{align}
\end{enumerate}
From these relations, we can identify possible constraints as
\begin{align}
\label{eq:vekconstr}    &\vek{}\mathcal{A}=2c_1+c_2+c_3=0 & \nonumber\\
    \implies &\vek C_B= \frac{\kappa}{2 \phi \sqrt{\gamma}} \normalvector^A \pi_{AB} + \frac{c_3}{2} T^Q{}_{PD} \pr^D_B \pr^P_Q\approx 0\,,\\
\label{eq:asyconstr}    &\asy{}\mathcal{A}=2c_1-c_2=0 \nonumber\\
    \implies &\asy C_{AB} =\frac{1}{2}\pr^C_{[A} \pr^D_{B]} \left(\frac{\kappa}{ \phi \sqrt{\gamma}}\pi_{CD} - c_2\normalvector^E T_{ECD}\right)\approx 0\,,\\
\label{eq:syconstr}    &\sym{}\mathcal{A}=2c_1+c_2=0 \nonumber\\
    \implies &\sym C_{AB} =
    \frac{\kappa}{2\phi\sqrt{\inducedmetric}}\left(\pr^C_{(A} \pr^D_{B)} - \frac{1}{3}(\eta_{AB} - \normalvector_A \normalvector_B) \eta^{CD}\right)\pi_{CD}\approx 0 \,,\\
\label{eq:trconstr}    &\trc{} \mathcal{A}=2c_1+c_2+3c_3=0 \nonumber\\
    \implies &\trc C = \frac{\kappa}{2 \phi \sqrt{\gamma}} \pi^A{}_A \approx 0\,.
\end{align}

The choice of values for the parameters $c_1, c_2, c_3$ determines which of the relations ${}^{\mathcal{I}}C = 0, \  \mathcal{I}=\mathcal{V,A,S,T}$ constitutes a primary constraint on the canonical variables. Setting the scalar field $\phi=1$ in these equations, one obtains the corresponding relations for the NGR class of theories.

For the $f(\mathbb{T})$ choice \eqref{eq:NGRTEGR}, the vector and antisymmetric parts impose primary constraints, while the trace and the symmetric trace free parts represent invertible equations which relate the time derivatives of the tetrad to the momenta. The general analysis for all $f(\tngr)$ classes gives the following additional constraints to the universal ones \eqref{eq:primfngr}:
\begin{table}[htbp!]
\begin{center}
	\begin{tabular}{|c|c|c|c|c|c|}
		\hline
		\multicolumn{1}{|c|}{} & \multicolumn{4}{|c|}{Theory parameter combinations} & \multicolumn{1}{|c|}{\#} \\ \hline
        ${}^{\mathcal{I}}\mathcal{A}$ & $2c_1 + c_2 + c_3$ & $2c_1 - c_2$ & $2c_1 + c_2$ & $2c_1 + c_2 + 3 c_3$ & \\ \hline
        \begin{tabular}{@{}c@{}} Constraints \\ if ${}^\mathcal{I}\mathcal{A}=0$ \end{tabular} & $\vek C_B\approx 0$ & $\asy C_{AB}\approx 0$ & $\sym C_{AB}\approx 0$ & $\trc C \approx 0$ & \\ \hline
        Case 1 & $\neq 0$ & $\neq 0$ & $\neq 0$ & $\neq 0$ & $0$ \\ \hline
        Case 2 & $0$ & $\neq 0$ & $\neq 0$ & $\neq 0$ & $3$ \\ \hline
        Case 3 & $\neq 0$ & 0 & $\neq 0$ & $\neq 0$ & $3$ \\ \hline
        Case 4 & $\neq 0$ & $\neq 0$ & $0$ & $\neq 0$ & $5$  \\ \hline
        Case 5 & $\neq 0$ & $\neq 0$ & $\neq 0$ & $0$ & $1$ \\ \hline
        Case 6 & $0$ & 0 & $\neq 0$ & $\neq 0$ & $6$ \\ \hline
        Case 7 & $\neq 0$ & $0$ & $0$ & $\neq0$ & $8$ \\ \hline
        Case 8 & $\neq 0$ & $0$ & $\neq 0$ & $0$ & $4$\\ \hline
        Case 9 & $ 0$ & $\neq 0$ & $ 0$ & $0$ & $9$ \\ \hline
	\end{tabular}
\end{center}
\caption{All possible non-trivial combinations of primary constraints in $f(\tngr)$. The last column denoted \# contains the number of independent primary constraints which are incurred in case that the given combinations of theory parameters vanish.}
\label{table:NGRconstraints}
\end{table}
No more possibilities to set combinations of the parameters $c_1, c_2$ and $c_3$ to zero without fixing all of them to zero exist. This classifies all possible primary constraints in the~$f(\tngr)$ class of theories (Table \ref{table:NGRconstraints}).

As a final remark of this section, we would like to display the constraints for the class of theories defined by the relation $(2c_1 + c_2 + c_3)=(2c_1 - c_2)=0$, which contains basically TEGR and $f(\mathbb{T})$ gravity, in the following compact form:
\begin{align}\label{eq:Lorentz}
      C_{AB} =   \pi_{[AB]} - \frac{1}{\kappa} \phi \sqrt{\gamma} n^D S_{[A|D|B]} \approx 0.
\end{align}
In TEGR ($\phi=1$), these constraints represent the freedom of applying local Lorentz transformations only to the tetrad alone without considering a spin connection. An important remark is that such constraints slightly differ from the Lorentz constraints obtained in the tetrad formulation of GR \cite{Deser:1976ay,Castellani:1981ue}. \footnote{However, it is possible to consider tetrad-based formulation of GR in the same sense as it has been done for TEGR. This was considered in \cite{NESTER1989112}.} Lorentz constraints from tetradic GR and TEGR are different since in TEGR there is an additional term, represented by the second term depending on the superpotential $S$ in \eqref{eq:Lorentz}. This essential difference has not been noticed enough in the literature (probably only mentioned in \cite{Ferraro:2016wht} and \cite{Ferraro:2020tqk}), but it can be understood by considering that the TEGR Lagrangian is pseudo-invariant under local Lorentz transformations. That is, the Lagrangian is modified by a four-divergence once we perform such transformations, which is integrated out once in the action. This fact has a great importance for $f(\mathbb T)$ gravity, where the four-divergence is not integrated out and the Lorentz symmetry of the tetrads alone
is partially or totally broken \cite{Ferraro:2020tqk}.

%
%
%
%
%------------------------------------------Dictionary---------------------------------------
%
%
%
%
%

\section{Dictionary relating the analysis from different authors}
\label{sec:Dictionary}
Using the general form of the primary constraints in $f(\tngr)$ theories introduced in the previous section, we can now relate the primary constraints obtained by other approaches and the more specific $f(\mathbb{T})$ and NGR classes of theories. We briefly summarize the key aspects of the discussed approaches in Sec.~\ref{ssec:summary}. A detailed discussion is then given in Sec.~\ref{ssec:dictionary}.

%
%
%
%
%---------------------------------------Summary of methods
%
%
%
%
%

\subsection{Summary of methods used by different authors}
\label{ssec:summary}

\begin{table}[htpb!]
\begin{center}
		{\begin{tabular}{@{}cccccc@{}} \toprule
		&  Blagojević &  Blixt & Ferraro & Maluf & Okołów \\ \colrule
		TEGR & \cite{Blagojevic:2000qs}  & - & \cite{Ferraro:2016wht} & \cite{Maluf:2001rg,Maluf:2013gaa} & \cite{Okolow:2011nq,Okolow:2013lwa} \\
		$f(\mathbb{T})$ & \cite{Blagojevic:2020dyq} & - & \cite{Ferraro:2018tpu} &  \cite{Li:2011rn}\footnote{This reference essentially uses the same notation as Maluf} & - \\
		NGR & \cite{Mitric:2019rop}\footnote{This reference essentially uses the same notation as Blagojevic} & \cite{Blixt:2018znp} & - & - & \cite{Okolow:2011np} \\ \colrule
		Spacetime indices & $\mu,\nu, \rho\ldots$ & $\mu,\nu,\rho\ldots$ & $\mu,\nu, \rho\ldots$ & $\mu,\nu, \rho\ldots$ & $\mu,\nu,\rho\ldots$ \\ \colrule
		Lorentz indices & $i,j,k\ldots$ & $A,B,C\ldots$ & $a,b,c\ldots$ & $a,b,c\ldots$ & $A,B,C\ldots$ \\ \colrule
		Time index & 0 & 0 & 0 & 0 & 0 \\ \colrule
		Spatial indices & - & $i,j,k\ldots$ & $i,j,k\ldots$ & $i,j,k\ldots$ & $i,j,k\ldots$ \\ \colrule
		Tetrad & $\vartheta^i$ & $\theta^A$  & $E^a$ & $e^a$ & $\boldsymbol{\theta}^A$\\ \colrule
		Cotetrad & $e_i$ & $e_A$ & $e_a$ & $e_a$ & - \\ \colrule
		$\det$ of tetrad & $\vartheta$ & $|\theta|$ & $E$ & $e$ & - \\ \colrule
		Metric sign $\eta_{00}$	& $+1$ & $-1$  & $+1$ & $-1$ & $-1$ \\ \colrule
		Lorentz 3+1	& Yes & No & No & No & No \\ \colrule
		Spin connection & $\omega^{ij}$ & $\omega^A{}_{B}$ & 0 & 0 & 0 \\ \colrule
		Tetrad momenta & $\pi_i{}^\mu$ & $\pi_A{}^i$ \footnote{For the auxiliary field related to the spin connection the notion of its conjugate momenta is $\hat{\pi}^{AB}$} & $\Pi^{\mu}_a$ & $\Pi^{a\mu}$ & $p_A$ \\ \colrule
		Scalar momenta & $\pi_\phi$ & - & $\pi$ & - & - \\ \colrule
		NGR Coefficients & $h_1,h_2,h_3$ & $c_1,c_2,c_3$ & - & $A,B,C$ & $a_1,a_2,a_3$ \\ \colrule
		Lapse & $N$ & $\alpha$ & - & - & $N$ \\ \colrule
		Shift & $N^\alpha$ & $\beta^i$ & - & - & $\vec{N}$ \\ \colrule
		Induced metric & implicit & $h_{ij}$ & - & - & $q_{ij}$ \\ \colrule
		Normal vector & $n_i$ & $\normalvector^A$ & - & - & $\xi^A$ \\ \botrule
		\end{tabular}}
		\end{center}
		\caption{Dictionary for notation of various authors (only the first author is given).}
			\label{dtable}
	\end{table}

In Table \ref{dtable} we summarize the notation used by different authors in the Hamiltonian formalism for TEGR, regarding the use of indices and definition of fields. Some of the formalisms introduced here have been applied for NGR and $f(\mathbb{T})$ gravity cases. The heading of each column denotes the surname of the first one of the authors (in alphabetic order), which represents a groups of independent references that encompass similar notation and analysis. We enumerate the different groups as follows:
\begin{enumerate}
	\item \textbf{Blagojević et al.} (Sec.~\ref{sssec:blagojevic}): In \cite{Blagojevic:2000qs}, by M. Blagojević and I. A. Nikolic, the full Hamiltonian analysis for TEGR was performed. The Hamiltonian analysis and constraint algebra for NGR has been calculated in detail by P. Mitric at \cite{Mitric:2019rop}, but only for the case with the least amount of constraints. This analysis has been applied for $f(\mathbb{T})$ gravity in \cite{Blagojevic:2020dyq} by M. Blagojević and J. M. Nester.
	\item \textbf{Blixt et al.} (Sec.~\ref{sssec:blixt}): In \cite{Blixt:2018znp,Blixt:2019mkt}, by D. Blixt, M. Hohmann and C. Pfeifer, the Hamiltonian analysis for NGR has been introduced and primary constraints have been found for all possible cases. It was proved that the formalism is independent of the Weitzenb\"{o}ck choice in the connection. These two papers use a notation in tensor components; the same calculation is performed using differential form notation in~\cite{Hohmann:2019sys}, by M. Hohmann.
	\item \textbf{Ferraro et al.} (Sec.~\ref{sssec:ferraro}): In \cite{Ferraro:2016wht}, by R. Ferraro and M. J. Guzm\'an, the Hamiltonian formalism for TEGR has been introduced in a premetric approach. For $f(\mathbb{T})$ gravity, it has been studied in \cite{Ferraro:2018tpu}, some guidelines for $f(\mathbb T)$ in the Einstein frame in \cite{Ferraro:2018axk}, and the role of the pseudoinvariance in the Hamiltonian formalism in \cite{Ferraro:2020tqk}, by the same authors. The same notation has been used for the classification of primary constraints in NGR in \cite{Guzman:2020kgh} by M.J. Guzm\'an and Sh. Khaled-Ibraheem.
	\item \textbf{Maluf et al.} (Sec.~\ref{sssec:maluf}): There are several works of J. W. Maluf, J. F. da Rocha-Neto, A. A. Sousa and S. C. Ulhoa. on the Hamiltonian formalism of TEGR, some of them are: \cite{Maluf:1994ji} Hamiltonian analysis of TEGR with gauge fixing conditions, with a time gauge \cite{Maluf:2000ah}, considering a null surface and no time gauge condition \cite{Maluf:1998ae}, without any condition of gauge fixing, \cite{Maluf:2001rg} analysis of unimodular teleparallel gravity \cite{daRochaNeto:2011ir}, and a review on TEGR which includes a summary on Hamiltonian formalism of TEGR \cite{Maluf:2013gaa}. The formalism introduced by J. W. Maluf et al. has been used for $f(\mathbb{T})$ gravity in the work of \cite{Li:2011rn} by M. Li, R. X. Miao and Y. G. Miao.
	\item \textbf{Okołów} (Sec.~\ref{sssec:okolow}): Calculations are performed using the language of differential forms. In~\cite{Okolow:2011np} by A. Okol\'ow, a simple subcase of NGR is studied, where ${}^\mathcal{I}\mathcal{A}\neq 0$ for all $\mathcal{I}$ so that none of the primary constraints in Eqs. \eqref{eq:vekconstr}-\eqref{eq:trconstr}. TEGR is considered in the papers~\cite{Okolow:2011nq,Okolow:2013lwa}, by the same author. The Hamiltonian, primary constraints and their PB are calculated for all cases.
\end{enumerate}

%
%
%
%
%--------------------------------Dictionary subsection--------------------------------
%
%
%
%
%

	%
	%
	%
	%
	%-------------------------------------Dictionary: Primary constraints---------------------------
	%
	%
	%
	%

\subsection{Dictionary of primary constraints}
\label{ssec:dictionary}
We now provide a detailed discussion and comparison of the primary constraints derived by different authors. These are Blagojević et al. in Sec.~\ref{sssec:blagojevic}, Blixt et al. in Sec.~\ref{sssec:blixt}, Ferraro et al. in Sec.~\ref{sssec:ferraro}, Maluf et al. in Sec.~\ref{sssec:maluf} and Okołów et al. in Sec.~\ref{sssec:okolow}.
%
%
%
%
%--------------------------------------Blagojevic--------------------------------------
%
%
%
%
%

\subsubsection{Blagojević}
\label{sssec:blagojevic}

In \cite{Blagojevic:2000qs} the Hamiltonian analysis of new general relativity, \cite{PhysRevD.19.3524}, has been presented. In \cite{Blagojevic:2000pi} the gauge symmetries of teleparallel gravity in the framework of Poincar\'{e} gauge gravity has been investigated. Finally, in \cite{Blagojevic:2020dyq} the Hamiltonian  analysis of $f(\mathbb{T})$ gravity, \cite{Ferraro:2006jd}, has been studied. The results of the investigations \cite{Blagojevic:2000qs}  and \cite{Blagojevic:2020dyq} are reproduced by our analysis in Sec. \ref{sec:fNGR} with help of the following relations.

First, for the analysis of the constraints of $f(\mathbb{T})$ gravity, the parameter choice \eqref{eq:NGRTEGR} has to be employed. Under this condition, the induction tensor \eqref{eq:induction} is directly related to the usual superpotential \eqref{eq:sp}, $H_{ABC} = \frac{1}{2}S_{ABC}$.

Second, in \cite{Blagojevic:2000qs} and \cite{Blagojevic:2020dyq} the following notation for the identification of the constraints from the irreducible decomposition \eqref{eq:vecfngr} to \eqref{eq:symfngr} is employed. In the local cotetrad basis, the normal vector to the hypersurfaces of the foliation of spacetime can be expanded as $\normalvector = \normalvector^A \cotetrad_A$. This can be used to expand every vector $V$ as
\begin{align}
    V = V^A \cotetrad_A  = V_\perp + \bar{V}\,,
\end{align}
where, with help of the projector $\pr^A_B = (\delta^A_B - n_B n^A )$, we can write
\begin{align}
    V_\perp = V^A \normalvector_A\ n,\quad \bar V = V -  V_\perp  = V^B \pr^A_B e_B = V^{\bar A} e_A\,,
\end{align}
i.e.\ barred indices are projected indices. Moreover $\sqrt{\gamma} = \frac{\dettetrad}{\alpha} = J$ and $a_0 = \frac{1}{2\kappa}$. In addition, in \cite{Blagojevic:2000qs} and \cite{Blagojevic:2020dyq}, the momenta with lower Lorentz indices are denoted by $\hat \pi_{AB}$.

With this translation, it is easy to see the relations \eqref{eq:vecfngr} to \eqref{eq:trfngr} become
\begin{align}
    \normalvector^A\hat \pi_{AB}
    &=  - 2 a_0 \phi J  T^{\bar{A}}{}_{\bar{A}\bar B}\,,\label{eq:blag1}\\
    \hat \pi_{[\bar A \bar B]}
    &=  a_0 \phi J T_{\perp\bar A \bar B}\,,\label{eq:blag2}\\
    (\hat\pi_{(\bar A \bar B)} - \frac{1}{3}(\eta_{AB} - \normalvector_A \normalvector_B)\hat\pi^C{}_C)
    &= 2 a_0 \phi J \left(T_{(\bar A |\perp| \bar B)} - \tfrac{(\eta_{AB} - \normalvector_A \normalvector_B)}{3} T^{\bar C}{}_{\perp \bar C}\right) \,.\label{eq:blag4}\\
    \hat \pi^A{}_A
    &= 4 a_0 \phi J T^{\bar A}{}_{ \bar A \perp}\,,\label{eq:blag3}
\end{align}

The six independent components of Eqs.\eqref{eq:blag1} and \eqref{eq:blag2} are primary constraints, since the right-hand side of these relations do not contain any time derivatives of the tetrad, while the six independent components of Eqs. \eqref{eq:blag3} and \eqref{eq:blag4} relate momenta and time derivatives of the tetrad. In a compact way the constraints can be captured with help of Eq. \eqref{eq:hatpi} or Eq. \eqref{eq:Lorentz} as
\begin{align}
\label{LorentzBN}
     C_{AB} := \hat \pi_{[AB]} - 4 a_0 \phi J H_{[A |\perp| B]} = \hat \pi_{[AB]} - 2 a_0 \phi J S_{[A |\perp| B]} = 0\,.
\end{align}

%
%
%
%
%--------------------------------------Blixt--------------------------------------
%
%
%
%
%

\subsubsection{Blixt}
\label{sssec:blixt}

In \cite{Blixt:2018znp} the primary constraints for new general relativity are expressed in the irreducible decomposition under the rotation group. This also goes for the conjugate momenta which are decomposed as:
\begin{align}
\label{eq:irreducemomenta}
    \momenta_A{}^i=\vek\momenta^i\normalvector_A+\asy\momenta^{ji}\inducedmetric_{kj}\tetrad_A{}^k+\sym\momenta^{ji}\inducedmetric_{kj}\tetrad_A{}^k+\trc\momenta \tetrad_A{}^i.
\end{align}
This leads to primary constraints consistent with Eqs. \eqref{eq:vecfngr}-\eqref{eq:trfngr}. They are expressed in spatial components instead of Lorentz components and explicitly written out in lapse and shift. Coupling to matter is not considered in their work and their coefficients $c_1, \ c_2, \ c_3$ differ from those in Eq. \eqref{eq:induction} by a factor $\frac{1}{2\kappa}$. Starting with the vector constraint satisfied in theories with $2c_1+c_2+c_3=0$ and $\phi=1$ which brings \eqref{eq:vecfngr} to
  \begin{align}
        \frac{\kappa}{2 \phi \sqrt{\gamma}} \normalvector^A\hat \pi_{AB}&= \frac{c_3}{2} T^Q{}_{PD} \pr^D_B \pr^P_Q\,.
    \end{align}
    Multiplying this equation with $\tetrad^B{}_j\inducedmetric^{ij}$ yields\footnote{Note that the relative sign of this constraint differs from \cite{Blixt:2018znp}. They have a different sign convention but Eq. \eqref{eq:irreducemomenta} is defined in the same way. When the irreducible vector part, then is contracted with $\normalvector^A$; this particular term will have the opposite sign compared to those in \cite{Blixt:2018znp}. Note that we could have defined $\vek \momenta^i$ with the opposite sign to make the expression look like the one in \cite{Blixt:2018znp}.}
    \begin{align}
        \begin{split}
            0&=\frac{\kappa}{2\sqrt{\inducedmetric}}\normalvector^A\momenta_A{}^\mu\tetrad^C{}_\mu \eta_{BC}\tetrad^B{}_j\inducedmetric^{ij}-\frac{c_3}{2}T^Q{}_{PB}\tetrad^B{}_j\inducedmetric^{ij}\left(\delta^P_Q-\normalvector_Q\normalvector^P \right) \\
            &=\frac{\kappa}{2\sqrt{\inducedmetric}}\normalvector^A\momenta_A{}^k\tetrad^C{}_k\eta_{BC}\tetrad^B{}_j\inducedmetric^{ij}-\frac{c_3}{2}T^P{}_{PB}\tetrad^B{}_j\inducedmetric^{ij}+\frac{c_3}{2}T^Q{}_{PB}\tetrad^B{}_j\inducedmetric^{ij}\normalvector_Q\normalvector^P\\
            &=\frac{\kappa}{2\sqrt{\inducedmetric}}\vek\momenta^i+\frac{c_3}{2}T^B{}_{kl}\tetrad_B{}^l\inducedmetric^{ik}
        \end{split}
    \end{align}
    The antisymmetric constraints are obtained for theories with $2c_1+c_2=0$ and $\phi=1$ which brings Eq. \eqref{eq:asymfngr} to
    \begin{align}
    \frac{\kappa}{2 \phi \sqrt{\gamma}}\pr^C_{[A} \pr^D_{B]}\hat\pi_{CD}\tetrad^A{}_i\tetrad^B{}_j
        &=  \frac{c_2}{2}\normalvector^D T_{DPQ}\pr^P_{[A}\pr^Q_{B]}\tetrad^A{}_m\tetrad^B{}_p\inducedmetric^{im}\inducedmetric^{jp}\,.
    \end{align}
     Further, we multiply this with $\tetrad^A{}_m\tetrad^B{}_p\inducedmetric^{im}\inducedmetric^{jp}$ so that
     \begin{align}
         \begin{split}
              0&=\frac{\kappa}{4\sqrt{\inducedmetric}}\momenta_A{}^k\tetrad^E{}_k\eta_{BE}\tetrad^A{}_m\tetrad^B{}_p\inducedmetric^{im}\inducedmetric^{jp}-\frac{\kappa}{4\sqrt{\inducedmetric}}\momenta_B{}^k\tetrad^E{}_k\eta_{AE}\tetrad^A{}_m\tetrad^B{}_p\inducedmetric^{im}\inducedmetric^{jp}\\
              &-\frac{c_2}{4}\normalvector_D\left(T^D{}_{AB}-T^D{}_{BA}\right)\tetrad^A{}_m\tetrad^B{}_p\inducedmetric^{im}\inducedmetric^{jp}\\
              &=\frac{\kappa}{2\sqrt{\inducedmetric}}\asy\momenta^{ji}-\frac{c_2}{2}\inducedmetric^{ik}\inducedmetric^{jl}T^B{}_{kl}\normalvector_B
         \end{split}
     \end{align}
    If the theory satisfies $2c_1-c_2=0$ and $\phi=1$, Eq. \eqref{eq:symfngr} reduces to
    \begin{align}
    \begin{split}
        \frac{\kappa}{2  \sqrt{\gamma}}    (\pr^C_{(A} \pr^D_{B)}\hat\pi_{CD} - \frac{1}{3}(\eta_{AB} - \normalvector_A \normalvector_B)\hat\pi^C{}_C)=0 \,.
    \end{split}
    \end{align}
    Similar to the antisymmetric constraints, we multiply this with $\tetrad^A{}_m\tetrad^B{}_p\inducedmetric^{im}\inducedmetric^{jp}$ and get
     \begin{align}
        \begin{split}
            0&=\frac{\kappa}{4\sqrt{\gamma}}\pi_A{}^k\theta^E{}_k\eta_{BE}\theta^A{}_m\theta^B{}_p\gamma^{im}\gamma^{jp}+\frac{\kappa}{4\sqrt{\gamma}}\pi_B{}^k\theta^E{}_k\eta_{AE}\theta^A{}_m\theta^B{}_p\gamma^{im}\gamma^{jp} \\
            & -\frac{\kappa}{6\sqrt{\gamma}}\pi_C{}^C\gamma^{ij}\\
            &=\frac{\kappa\pi^{(ji)}}{2\sqrt{\gamma}}-\frac{\kappa{}^\mathcal{T}pi}{2\sqrt{\gamma}}\gamma^{ij}\\
            &=\frac{\kappa}{2}\cdot\frac{{}^\mathcal{S}\pi^{ij}}{\sqrt{\gamma}}.
        \end{split}
    \end{align}
        If $2c_1+c_2+3c_3$ and $\phi=1$ are satisfied, Eq. \eqref{eq:trfngr} reads
    \begin{align}
        \frac{\kappa}{2 \phi \sqrt{\gamma}} \eta^{AB}\pr^C_{A} \pr^D_{B}\hat\pi_{CD}
        &= \frac{\kappa}{2 \phi \sqrt{\gamma}}\hat \pi^A{}_A = 0 \,.
    \end{align}
Expanding this notion from the definition gives
\begin{align}
    \begin{split}
        0&=\frac{\kappa}{2\sqrt{\inducedmetric}}\momenta_A{}^\mu \tetrad^C{}_\mu\eta^{AB}\eta_{BC}=\frac{\kappa}{2}\frac{\momenta_A{}^A}{\sqrt{\inducedmetric}}=\frac{\kappa}{2}\frac{\momenta_\mu{}^\mu}{\sqrt{\inducedmetric}}=\frac{\kappa}{2}\frac{\momenta_i{}^i}{\sqrt{\inducedmetric}}
        \\ &=\frac{\kappa}{2}\frac{\trc\momenta}{\sqrt{\inducedmetric}}.
    \end{split}
\end{align}
All of the above constraints can now easily be seen to be consistent with \cite{Blixt:2018znp}.

%
%
%
%
%-------------------------------Ferraro---------------------------------------------------
%
%
%
%
%

\subsubsection{Ferraro}
\label{sssec:ferraro}

We will present the primary constraints for $f(\mathbb T)$ gravity as presented in \cite{Ferraro:2018tpu}. The Jordan frame representation of $f(\mathbb T)$ gravity has been taken as a starting point, with the Lagrangian
\begin{align}
\label{LgrFG}
    L_{\mathrm{f}} =\dettetrad [ \phi \mathbb T - V(\phi) ].
\end{align}
This differs from \eqref{Lfngr} only in the gravitational constant factor $\frac{1}{2\kappa}$, which appears in the action instead of the Lagrangian. The torsion scalar is rewritten as
\begin{equation}
\mathbb T = \tetrad \partial_{\mu} \tetrad^{A}{}_{\nu} \partial_{\rho} \tetrad^{B}{}_{\lambda} \cotetrad_C{}^{\mu} \cotetrad_E{}^{\nu} \cotetrad_D{}^{\rho} \cotetrad_F{}^{\lambda} \chi_{AB}{}^{CEDF},
\end{equation}
where the object $\chi_{AB}{}^{CEDF}$ is the constitutive tensor, a mathematical object depending only on the components of the Minkowski metric and Kronecker deltas. Although in \cite{Ferraro:2018tpu} it was only considered the generalization of TEGR, it is possible to write this object for the most general NGR case as \cite{Guzman:2020kgh}
\begin{equation}
 \chi_{AB}{}^{CEDF} = 4 c_1 \eta_{AB} \eta^{C[D} \eta^{F]E} - 4 c_2 \delta_{A}^{[D}\eta^{F][C}\delta^{E]}_{B} + 4 c_3 \delta_{A}^{[C} \eta^{E][D}\delta^{F]}_{B},
\end{equation}
where the particular TEGR case is obtained for the values of the $c_i$ presented in Eq. \eqref{eq:NGRTEGR}.

The canonical momenta are defined as
\begin{equation}
\label{pFG}
\momenta_A{}^{\mu} = \dfrac{\partial L}{\partial \partial_0\tetrad^{A}{}_{\mu} } = \phi \tetrad \partial_{\rho} \tetrad^{B}{}_{\lambda} \cotetrad_C{}^0 \cotetrad_E{}^{\mu} \cotetrad_D{}^{\rho} \cotetrad_F{}^{\lambda} \chi_{AB}{}^{CEDF},
\end{equation}
where we remark that the ADM decomposition has not been used for the tetrad. The following trivial constraints appear
\begin{equation}
    C_A = \pi_A{}^{0} \approx 0.
\end{equation}
These can be extracted from \eqref{pFG} by noticing that for $\mu=0$, it appears the pair $\cotetrad_C{}^0 \cotetrad_E{}^{0}$ which is symmetric in $CE$, but is multiplied by the constitutive tensor $\chi_{AB}{}^{CEDF}$ which is antisymmetric on such indices. These constraints are equivalent to  \eqref{pcBHP}.

In Ref.\cite{Ferraro:2018tpu}, it has been proposed an alternative way of obtaining these primary constraints in terms of the kernel of the Hessian matrix $C_{AB}{}^{EF}=\cotetrad_C{}^{0} \cotetrad_D{}^{0}\chi_{AB}{}^{CEDF}$. The canonical momenta in  \eqref{pFG} are rewritten in terms of $C_{AB}{}^{EF}$ as
\begin{equation}
\momenta_{A}{}^{\mu} \tetrad^{E}{}_{\mu} = \tetrad C_{AB}{}^{EF} \cotetrad_F{}^{\lambda} \dot{\tetrad}\tetrad^{B}{}_{\lambda} + \tetrad \partial_i \tetrad^{B}{}_{\lambda} \cotetrad_C{}^{0} \cotetrad_D{}^{i} \cotetrad_{F}{}^{\lambda}  \chi_{AB}{}^{CEDF}.
\end{equation}
By noticing that $\cotetrad_E{}^0 \delta^{A}_{G}$ lies in the kernel of the Hessian, that is
\begin{equation}
\cotetrad_E{}^0  \delta^{A}_{G} C_{AB}{}^{EF} = \cotetrad_E{}^0 \cotetrad_C{}^0 \cotetrad_D{}^0 \delta^{A}_{G} \chi_{AB}{}^{CEDF} = 0,
\end{equation}
we obtain the primary constraints $C_A \approx 0$.
Notice that in Ref.\cite{Ferraro:2018tpu}, uppercase Latin indices have been used to define the superindices (which denote pairs of Lorentz indices). In order to avoid confusion, we have omitted their use in this review.

Lorentz constraints can be obtained by an additional set of vectors in the kernel given by $2\delta^{A}_{[G}\eta_{H]E}$. We can prove that these lie in the kernel by calculating
\begin{equation}
2\delta^{A}_{[G}\eta_{H]E} C_{AB}{}^{EF} = 2\delta^{A}_{[G}\eta_{H]E} \cotetrad_C{}^0 \cotetrad_D{}^0 \chi_{AB}{}^{CEDF}.
\end{equation}
We obtain the following contraction of the constitutive tensor:
\begin{equation}
\delta^{A}_{[G}\eta_{H]E} \chi_{AB}{}^{CEDF} = \eta_{E[H} \chi_{G]B}{}^{CEDF} = - 2\delta_{GHB}^{CDF}.
\end{equation}

The triple totally antisymmetrized  Kronecker delta $\delta^{GHF}_{CAB}$ has been defined in \cite[Eq. (A5)]{Ferraro:2016wht} as
\begin{equation}
-\delta^{GHF}_{CAB} = \delta^{H}_{[A} \delta^{F}_{C]} \delta^{G}_{B} + \delta^{G}_{[A} \delta^{H}_{C]} \delta^{F}_{B} + \delta^{F}_{[A} \delta^{G}_{C]} \delta^{H}_{B}.
\end{equation}
Here notice that the triple antisymmetrization has not been defined with the conventional $1/3$ factor. However, antisymmetrization of pairs of indices is defined as usual, that is
\begin{align}
    V_{[AB]}=\frac{1}{2}\left(V_{AB}-V_{BA} \right).
\end{align}
Taking into account these subtleties, it is found the primary constraints
\begin{equation}
	\label{LorentzFG}
	C_{AB} = 2\eta_{E[B}\momenta_{A]}{}^i \tetrad^{E}{}_i + 4 \tetrad \partial_i \tetrad^C{}_j \left( \cotetrad_{[B}{}^0 \cotetrad_{A]}{}^i \cotetrad_{C}{}^j + \cotetrad_{[B}{}^i \cotetrad_{A]}{}^j \cotetrad_{C}{}^0 + \cotetrad_{[B}{}^j \cotetrad_{A]}{}^0 \cotetrad_{C}{}^i \right).
\end{equation}

These constraints can alternatively be written as
\begin{equation}
C_{AB} = \pi_{AB} - \pi_{BA} - 2 \phi \tetrad   [ \tetrad_A{}^i \tetrad_B{}^j T^{0}{}_{ij} - ( \tetrad_A{}^{i} \tetrad_B{}^{0} - \tetrad_B{}^{i} \tetrad_A{}^{0} )T^{j}{}_{ij} ].
\end{equation}
This expression is found to be consistent with Eq. \eqref{eq:Lorentz}, as we will explicitly demonstrate in Sec. \ref{sssec:maluf}.

%
%
%
%
%--------------------------------------Maluf--------------------------------------
%
%
%
%
%

\subsubsection{Maluf}
\label{sssec:maluf}

In this section, we will present the primary constraints of the Hamiltonian formalism in $f(\mathbb T)$ gravity performed by Li, Miao, Miao \cite{Li:2011rn}. Their analysis heavily relies on the Hamiltonian analysis of TEGR done by W. Maluf \cite{Maluf:2013gaa}, so the name of this subsection. The authors \cite{Li:2011rn} start from the following action:
\begin{equation}
S = -\int d^{4}x \tetrad f(\mathbb T)
\end{equation}
where $G=\frac{1}{16\pi}$ was taken. After passing to the equivalent scalar-torsion form \eqref{Lfngr} and considering the form of the torsion scalar, $\mathbb T = T_{ABC} \Sigma^{ABC}=\frac{1}{2}T_{ABC}S^{ABC}$\footnote{In \cite{Maluf:2013gaa} there is an overall factor $k=\frac{1}{2 c \kappa}$, also the matter is rescaled compared to our convention by a factor $\frac{1}{c}$} identifies the $\Sigma^{ABC}$ as half the superpotential $S^{ABC}$. Note that in the convention of this paper, this action is recovered by choosing $\kappa=1$. We find that the momentum is
\begin{equation} \label{pmaluf}
\momenta^{A\mu}=\dfrac{\partial L}{\partial \partial_0 \cotetrad_{A\mu} } = - 4\phi \dettetrad\Sigma^{A0\mu},
\end{equation}
which is consistent with Eq. \eqref{eq:Lorentz}\footnote{Note that \cite{Li:2011rn} as well as \cite{Maluf:2013gaa} use the opposite metric sign convention $\eta_{\mu\nu}=\mathrm{diag}(-1,1,1,1)$ which must be taken into account when contracting $\normalvector^D$ in equation \eqref{eq:Lorentz}}. In this expression, notice that the first index in $\Sigma^{A0\mu}$ is Lorentzian and the next two are spacetime ones. Notice that it is considered $\Sigma^{ABC} = H^{ABC}$ like in \eqref{eq:induction}, but using the TEGR coefficients. After removing the trivial primary constraints $\momenta^{A0}\approx 0$ from \eqref{pmaluf}, and considering that the object $S^{A0i} = S^{ABC} \cotetrad_B{}^0 \cotetrad_C{}^i$ remains, one can obtain the following primary constraints:
\begin{equation}\label{climiaox2}
C^{AB} = \momenta^{AB}-\momenta^{BA} + 2\phi \dettetrad \left[ \tetrad^{Am} \tetrad^{Bj} T^{0}{}_{mj} - (\tetrad^{Am}\tetrad^{B0} - \tetrad^{Bm}\tetrad^{A0} ) T^{j}{}_{mj} \right]
\end{equation}
In the works of Maluf et al. \cite{Maluf:2013gaa} it can be found another form for these constraints (for TEGR, but they can be easily generalized for $f(\mathbb T)$): \footnote{Note that \cite{Maluf:2013gaa} has a different convention for antisymmetrization brackets compared to this article. Their convention is $\momenta^{[AB]}=\momenta^{AB}-\momenta^{BA}$.}
\begin{equation}
\label{LorentzM}
C^{AB} = - C^{BA} = \momenta^{[AB]} = \momenta^{AB}-\momenta^{BA} + 4k \phi \dettetrad(\Sigma^{A0B} - \Sigma^{B0A} ).
\end{equation}
We can prove that the term $\Sigma^{A0B} - \Sigma^{B0A}$ gives indeed the torsion components appearing in \eqref{climiaox2} as follows.
\begin{equation}\label{is0}
\Sigma^{A0B} - \Sigma^{B0A} = \dfrac12\left(T^{0}{}_{\mu\nu}\tetrad^{A\mu}\tetrad^{B\nu} - \tetrad^{A0}T^{\mu \ B}_{\ \mu} + \tetrad^{B0}T^{\mu\ A}_{\ \mu} \right)
\end{equation}
The first term in the previous expression corresponds to:
\begin{equation}\label{is1}
T^{0}{}_{\mu\nu}\tetrad^{A\mu}\tetrad^{B\nu} =  T^{0}{}_{0i}\tetrad^{Bi} \tetrad^{A0} + T^{0}{}_{i0} \tetrad^{B0} \tetrad^{Ai} + T^{0}{}_{ij}\tetrad^{Ai} \tetrad^{Bj},
\end{equation}
while the remaining terms can be worked as
\begin{eqnarray}\label{is2}
    - \tetrad^{A0}T^{\mu \ B}_{\ \mu} + \tetrad^{B0}T^{\mu\ A}_{\ \mu} & = & -\tetrad^{A0} \tetrad^{B\nu} T^{0}{}_{0\nu} + \tetrad^{B0} \tetrad^{A\nu} T^{0}{}_{0\nu} - \tetrad^{A0} \tetrad^{B\nu} T^{i}{}_{i\nu} + \tetrad^{B0}\tetrad^{A\nu} T^{i}{}_{i\nu} \nonumber \\
    & = & -\tetrad^{A0} \tetrad^{Bi} T^{0}{}_{0i} + \tetrad^{B0} \tetrad^{Ai} T^{0}{}_{0i} - \tetrad^{A0} \tetrad^{B0} T^{k}{}_{k0} - \tetrad^{A0} \tetrad^{Bi} T^{k}{}_{ki} \nonumber \\
    & & + \tetrad^{B0} \tetrad^{A0} T^{k}{}_{k0} + \tetrad^{B0} \tetrad^{Ai} T^{k}{}_{ki}.
\end{eqnarray}
By combining \eqref{is1} and \eqref{is2} in \eqref{is0}, we recover the form \eqref{climiaox2} for the Lorentz constraints.

%
%
%
%
%--------------------------------------Okolow--------------------------------------
%
%
%
%
%

\subsubsection{Okołów}
\label{sssec:okolow}

The action is written in the form
\begin{equation}
S = -\frac{1}{2}\int_{\mathcal{M}}\dd\tetrad^A \wedge \star\left(\sum_{i = 1}^3a_i\,\dd\tetrad_A^{(i)}\right) = \int_{\mathcal{M}}\dd t \wedge L_{\perp}\,,
\end{equation}
where \(\star\) is the Hodge star of the spacetime metric \(\metric\), \(\mathcal{M} = \mathbb{R} \times \Sigma\) is the spacetime manifold and \(L_{\perp}\) is a differential three-form. It depends on three constants \(a_{1,2,3}\) which determine a particular choice of the NGR Lagrangian. The three terms appearing in the action are the tensor, vector and axial torsion components
\begin{subequations}
\begin{align}
\dd\tetrad_A^{(1)} &= \dd\tetrad_A - \dd\tetrad_A^{(2)} - \dd\tetrad_A^{(3)}\,,\\
\dd\tetrad_A^{(2)} &= \frac{1}{3}\tetrad_A \wedge (\cotetrad_B \intprod \dd\tetrad^B)\,,\\
\dd\tetrad_A^{(3)} &= \frac{1}{3}\cotetrad_A \intprod (\tetrad_B \wedge \dd\tetrad^B)\,,
\end{align}
\end{subequations}
where the Weitzenböck gauge is assumed, so that no spin connection appears. Writing the torsion in the tetrad basis in the form
\begin{equation}
\dd\tetrad_A = T_A = \frac{1}{2}T_{ABC}\tetrad^B \wedge \tetrad^C\,,
\end{equation}
the three terms are related to the torsion components by
\begin{subequations}
\begin{align}
\dd\tetrad_A^{(1)} &= \frac{1}{2}\left(T_{ABC} - \frac{2}{3}\eta_{AB}T^D{}_{DC} - T_{[ABC]}\right)\tetrad^B \wedge \tetrad^C\,,\\
\dd\tetrad_A^{(2)} &= \frac{1}{3}\eta_{AB}T^D{}_{DC}\tetrad^B \wedge \tetrad^C\,,\\
\dd\tetrad_A^{(3)} &= \frac{1}{2}T_{[ABC]}\tetrad^B \wedge \tetrad^C\,.
\end{align}
\end{subequations}
Further using the expressions for the Hodge star given by
\begin{subequations}
\begin{align}
\star 1 &= \frac{1}{4!}\epsilon_{ABCD}\tetrad^A \wedge \tetrad^B \wedge \tetrad^C \wedge \tetrad^D = \dettetrad\,\dd^4x\,,\\
\star(\tetrad^A \wedge \tetrad^B) &= \frac{1}{2}\epsilon^{ABCD}\tetrad_C \wedge \tetrad_D\,,
\end{align}
\end{subequations}
from which follows the inner product
\begin{equation}
\tetrad^A \wedge \tetrad^B \wedge \star(\tetrad^C \wedge \tetrad^D) = 2\eta^{A[C}\eta^{D]B} \star 1\,,
\end{equation}
the three terms in the action are found to be
\begin{subequations}
\begin{align}
\dd\tetrad^A \wedge \star\dd\tetrad_A^{(1)} &= \frac{1}{3}(T^{ABC}T_{ABC} + T^{ABC}T_{CBA} - T_A{}^{AC}T^B{}_{BC}) \star 1\,,\\
\dd\tetrad^A \wedge \star\dd\tetrad_A^{(2)} &= \frac{1}{3}T_A{}^{AC}T^B{}_{BC} \star 1\,,\\
\dd\tetrad^A \wedge \star\dd\tetrad_A^{(3)} &= \frac{1}{6}T^{ABC}(T_{ABC} - 2T_{CBA}) \star 1\,.
\end{align}
\end{subequations}
Hence, by comparing with the general Lagrangian~\eqref{eq:induction}, one finds that the constants \(a_{1,2,3}\) are related to \(c_{1,2,3}\) by
\begin{equation}
c_1 = \frac{1}{6}(2a_1 + a_3)\,, \quad
c_2 = \frac{1}{3}(a_1 - a_3)\,, \quad
c_3 = \frac{1}{3}(a_2 - a_1)\,,
\end{equation}
or equivalently,
\begin{equation}
a_1 = 2c_1 + c_2\,, \quad
a_2 = 2c_1 + c_2 + 3c_3\,, \quad
a_3 = 2c_1 - 2c_2\,,
\end{equation}
where in addition the convention \(\kappa \equiv -1\) for the value of the gravitational constant and the sign of the action are used. Setting \(a_1 = a_2 = a_3 = 1\) then yields the toy model studied in~\cite{Okolow:2011np}, while for \(a_1 = 1, a_2 = -2, a_3 = -1/2\), one obtains TEGR~\cite{Okolow:2011nq,Okolow:2013lwa}. They are related to the constants in~\cite{Hohmann:2019sys} by \(a_1 = -2C_T, a_2 = -2C_V, a_3 = -2C_A\).

The tetrad one-form is then split in the form
\begin{equation}
\tetrad^A = \tetrad^A{}_{\mu}\dd x^{\mu} = \tetrad^A{}_0\dd t + \tetrad^A{}_i\dd x^i = \tetrad^A_{\perp}\dd t + \vec{\tetrad}^A\,.
\end{equation}
Time derivatives are defined as
\begin{equation}
\dot{\tetrad}^A = \mathcal{L}_{\partial_t}\tetrad^A\,.
\end{equation}
One finds that only time derivatives of the spatial tetrad components \(\vec{\tetrad}^A\) appear in the Lagrangian \(L_{\perp}\), but not of the time components \(\tetrad^A_{\perp}\). Thus, the canonical momenta are introduced only for the spatial tetrad components, since they would be vanishing for the time components. They are defined as the differential two-forms \(\momenta_A\) by
\begin{equation}
\delta_{\dot{\tetrad}}L_{\perp} = \delta\dot{\vec{\tetrad}}^A \wedge \momenta_A\,,
\end{equation}
which are related to the momenta \(\momenta_A{}^{\mu}\) in the definition~\eqref{eq:pifNGR} by
\begin{equation}
\momenta_A{}^{\mu}\,\dd^4x = \dd t \wedge \dd x^{\mu} \wedge \momenta_A\,.
\end{equation}
Explicitly, they are given by
\begin{multline}
\momenta_A = \frac{1}{3\lapse}\Big\{(2a_1 + a_2)\ast\left[\dot{\vec{\tetrad}}_A - \dd(\lapse\normalvector_A) - \mathcal{L}_{\vec{\shift}}\vec{\tetrad}_A\right]\\
+ (a_1 - a_2)\vec{\tetrad}_B \wedge \ast\left(\dot{\vec{\tetrad}}^B \wedge \vec{\tetrad}_A + E^B{}_A\right) + (a_3 - a_1)\vec{\tetrad}_A \wedge \ast\left(\dot{\vec{\tetrad}}^B \wedge \vec{\tetrad}_B + E^B{}_B\right)\Big\}
\end{multline}
for general NGR~\cite{Hohmann:2019sys}, using the Hodge star \(\ast\) of the induced metric \(\inducedmetric\) on the spatial hypersurfaces, as well as the abbreviation
\begin{equation}
E^B{}_A = -\dd(\lapse\normalvector^B) \wedge \vec{\tetrad}_A + \lapse\normalvector_A\dd\vec{\tetrad}^B - (\mathcal{L}_{\vec{\shift}}\vec{\tetrad}^B) \wedge \vec{\tetrad}_A\,.
\end{equation}
Rewriting this in the parameters \(c_{1,2,3}\), we thus have
\begin{multline}
\momenta_A = \frac{1}{\lapse}\Big\{(2c_1 + c_2 + c_3)\ast\left[\dot{\vec{\tetrad}}_A - \dd(\lapse\normalvector_A) - \mathcal{L}_{\vec{\shift}}\vec{\tetrad}_A\right]\\
- c_3\vec{\tetrad}_B \wedge \ast\left(\dot{\vec{\tetrad}}^B \wedge \vec{\tetrad}_A + E^B{}_A\right) - c_2\vec{\tetrad}_A \wedge \ast\left(\dot{\vec{\tetrad}}^B \wedge \vec{\tetrad}_B + E^B{}_B\right)\Big\}
\end{multline}
With the particular choices of the three constant parameters one thus obtains
\begin{equation}
\momenta_A = \frac{1}{\lapse}\ast\left[\dot{\vec{\tetrad}}_A - \dd(\lapse\normalvector_A) - \mathcal{L}_{\vec{\shift}}\vec{\tetrad}_A\right]
\end{equation}
for the toy model studied in~\cite{Okolow:2011np}, as well as
\begin{equation}
\momenta_A = \frac{1}{\lapse}\left[\vec{\tetrad}_B \wedge \ast\left(\dot{\vec{\tetrad}}^B \wedge \vec{\tetrad}_A + E^B{}_A\right) - \frac{1}{2}\vec{\tetrad}_A \wedge \ast\left(\dot{\vec{\tetrad}}^B \wedge \vec{\tetrad}_B + E^B{}_B\right)\right]
\end{equation}
for TEGR~\cite{Okolow:2011nq,Okolow:2013lwa}. Depending on the choice of the constant parameters \(a_i\), the following primary constraints may appear, which are given by
\begin{subequations}
\begin{align}
\vek{C}_A &= \ast\vek{\momenta}_A + \frac{1}{3}(a_1 - a_2)\normalvector_A\vec{\tetrad}_B^{\sharp} \intprod \dd\vec{\tetrad}^B\,,\\
\asy{C}_A &= \ast\asy{\momenta}_A + \frac{1}{3}(a_3 - a_1)\vec{\tetrad}_A^{\sharp} \intprod (\vec{\tetrad}^B \wedge \dd\normalvector_B)\,,\\
\sym{C}_A &= \ast\sym{\momenta}_A\,,\\
\trc{C}_A &= \ast\trc{\momenta}_A\,,
\end{align}
\end{subequations}
where we made use of the musical isomorphism, which for a one-form \(\tau_A\) yields the vector field
\begin{equation}
\tau_A^{\sharp} = \inducedmetric^{-1}(\cdot, \tau_A)
\end{equation}
and the irreducible decomposition introduced in section~\ref{sec:fNGR}, which acts on one-forms as
\begin{subequations}
\begin{align}
\vek{\tau}_A &= -\normalvector_A\normalvector^B\tau_B\,,\\
\asy{\tau}_A &= \frac{1}{2}\left[\inducedmetric^{-1}(\vec{\tetrad}_A, \vec{\tetrad}^B)\tau_B - \inducedmetric^{-1}(\vec{\tetrad}_A, \tau_B)\vec{\tetrad}^B\right]\,,\\
\sym{\tau}_A &= \frac{1}{2}\left[\inducedmetric^{-1}(\vec{\tetrad}_A, \vec{\tetrad}^B)\tau_B + \inducedmetric^{-1}(\vec{\tetrad}_A, \tau_B)\vec{\tetrad}^B\right] - \frac{1}{3}\inducedmetric^{-1}(\vec{\tetrad}^B, \tau_B)\vec{\tetrad}_A\,,\\
\trc{\tau}_A &= \frac{1}{3}\inducedmetric^{-1}(\vec{\tetrad}^B, \tau_B)\vec{\tetrad}_A\,,
\end{align}
\end{subequations}
is extended to the momentum two-forms as \(\ast\prescript{\bullet}{}{\pi_A} = \prescript{\bullet}{}{(\ast\pi_A)}\).

%
%
%
%
%----------------------------------------Discussion-----------------------------------------------
%
%
%
%
%

\section{Discussion}
\label{sec:Discussion}

An essential point in the discussion of the Hamiltonian formalism of modified teleparallel gravities is the correct implementation of the Dirac-Bergmann algorithm. Two crucial steps in the algorithm that should be taken with care are: (i) that the Hessian could have variable rank, which we discuss in Sec.~\ref{ssec:hessian}, and (ii) that the matrix of PB among constraints has variable rank, as discussed in Sec.~\ref{ssec:pbmatrix}. An important conclusion one can draw from the Hamiltonian analysis is the number of degrees of freedom which, when compared with the outcome of perturbation theory, may reveal eventual strongly coupled fields; this will be discussed in Sec.~\ref{ssec:perttheory}.

\subsection{The Hessian in modified teleparallel gravities}
\label{ssec:hessian}
A caveat of concern in the proper application of the Dirac-Bergmann algorithm occurs when we are in presence of a Hessian that can have variable rank once evaluated in the constraint surface \cite{PhysRevA.54.57}. The Hessian for our classical finite-dimensional system
\begin{align}
    W_{ij}=\frac{\partial^2 L(q^{k},\dot{q}^{k})}{\partial \dot{q}^i \partial\dot{q}^j},
\end{align}
can be generalized for a field theory dependent on the tetrad as the following tensor:
\begin{align}
W_{AB}{}^{\mu\nu} = \frac{\partial^2 L(\tetrad^{C}{}_{\lambda}, \dot{\tetrad}^{C}{}_{\lambda}) }{\partial \partial_0 \tetrad^{A}{}_{\mu} \partial \partial_0 \tetrad^{B}{}_{\nu}  },
\end{align}
whose expression has mixed Lorentz and spacetime indices.
The full Hessian for NGR was presented in \cite{Guzman:2020kgh} as
\begin{equation}\label{Hessngr}
W_{AB}{}^{\mu\nu} = \tetrad \cotetrad_C{}^{0} \cotetrad_D{}^{0} \cotetrad_E{}^{\mu} \cotetrad_F{}^{\nu} \chi_{AB}{}^{CEDF} = \cotetrad_E{}^{\mu} \cotetrad_F{}^{\nu} \tilde{W}_{AB}{}^{EF},
\end{equation}
where the expression $\tilde{W}_{AB}{}^{EF}=\cotetrad_C{}^{0} \cotetrad_D{}^{0} \chi_{AB}{}^{CEDF}$ has been explicitly written in matrix form in \cite[Appendix A]{Guzman:2020kgh}. We notice that the NGR Hessian is linearly dependent on a quartic combination of cotetrad components $\cotetrad_A{}^{\mu}$. It is reasonable to assume that in the NGR case such expression does not vanish on the constraint surface. This is because no dynamical terms for the tetrad or cotetrad appear there; therefore, no constraints can show up. Therefore, the rank of the Hessian in NGR remains constant.

We note that the Hessian~\eqref{Hessngr} for NGR can be easily extended to the $f(\tngr)$ gravity case by following Eq. \eqref{Lfngr} and realizing that the $f(\tngr)$ Hessian is just the NGR Hessian multiplied by the auxiliary scalar field $\phi$. For $f(\tngr)$, the fundamental fields are the tetrads and the scalar field $\phi$, however, the Lagrangian does not contain any derivatives of $\phi$. Hence, the components of the Hessian emerging through the extra scalar field are zero, since  $\partial_{\dot \phi}L = 0$. Therefore, the nonvanishing components of the $f(\tngr)$ Hessian are obtained from \eqref{Hessngr} by multiplication with $\phi$.

Note that $\phi$ is a field whose value in principle can be zero. This can impose a vanishing Hessian for $f(\tngr)$ as well, which would also be the case for $f(\mathbb{T})$ gravity. This unlikely case might need special considerations in the Dirac-Bergmann algorithm \cite{PhysRevA.54.57}.

\subsection{Matrix of PB among constraints with variable rank}
\label{ssec:pbmatrix}

There are several indications that modified teleparallel theories suffer from variable rank on their matrix of PB among constraints. First, it was found that in the one-parameter teleparallel gravity model \cite{Cheng:1988zg} a field-dependent PB among constraints exists, which would change its value for certain tetrad configurations. The authors suggest that this could be a generic feature of teleparallel theories. Later, for Poincar\'e gauge gravity similar results were found \cite{Chen:1998ad,Yo:1999ex,Yo:2001sy,Barker:2021oez}. For $f(\mathbb T)$ gravity, such findings appear throughout the literature but probably their impact has not been stressed enough, as for instance in \cite{Li:2011rn,Ferraro:2018tpu,Blagojevic:2020dyq}. There is an ongoing discussion in the literature regarding the physical number of degrees of freedom in $f(\mathbb T)$, and although a better understanding on the constraint structure is needed, it is believed that in the most general case, there are 5 degrees of freedom. A controversial point is what happens when the scalar field $\phi$ does not depend on the spatial hypersurface coordinates, for example $\phi=\phi(x^0)$ only. In this case, the PB of the Lorentz constraints is weakly zero, and there is some apparent recovery of the invariance of the theory under pure tetrad Lorentz transformations. However, there could still be room for other pieces of the algebra to be different from zero, since some Lorentz constraints could  become second class due to noncommuting PB with additional primary constraints  \cite{Ferraro:2018tpu,Ferraro:2018axk}. Other evidence pointing toward 3 d.o.f. in this subcase comes from perturbative analysis \cite{Jimenez:2020ofm,Golovnev:2020nln}. The results do not immediately contradict each other, since more than one jump in the rank of the PB matrix among constraints could be possible. Further work to clarify the aforementioned discussion is motivated in order to resolve controversial points stated in the literature for $f(\mathbb T)$.

\subsection{Possible conclusions from perturbation theory}
\label{ssec:perttheory}

Although the Hamiltonian analysis provides solutions for the complete non-linear theory, perturbation theory becomes very important not only for doing practical calculation, but also for understanding the properties of the fundamental physical fields. In particular, applying perturbation theory around different backgrounds might reveal an issue with strongly coupled fields, which invalidates the perturbation theory around this particular background. Another important piece of information perturbation theory might give us is a consistency check of the Hamiltonian analysis. We do not expect more modes at the perturbative level than at the nonlinear level.

To summarize the work that has been done for the main teleparallel gravity theories discussed in the literature (that is TEGR, $f(\mathbb{T})$ and NGR), we collect results for perturbations around several backgrounds. At the end of the section, we list the main conclusions in a compact way first for linear perturbations and then for higher order.

Starting with TEGR, the second-order (lowest order) perturbations around a Minkowski background is consistent with linearized general relativity, see Ref. \cite{Ortin:2015hya} in Sec. 4.6. These perturbations are known to be those of a massless spin-2 field and consist of 2 degrees of freedom. Furthermore, many works agree, through a Hamiltonian analysis, that the full nonlinear theory propagates 2 degrees of freedom \cite{Blagojevic:2000qs,Maluf:2001rg,Ferraro:2016wht}.
Hence, it is expected that exactly 2 degrees of freedom propagate for TEGR in any background, which is explicitly observed for Minkowski and flat FLRW backgrounds\cite{Golovnev:2018wbh}.

For $f(\mathbb{T})$, the lowest order perturbations are very reminiscent of those of TEGR around diagonal tetrads for Minkowski spacetime \cite{Chen:2010va,Izumi:2012qj,Golovnev:2018wbh,Golovnev:2020aon} and, hence,  two modes propagate around these backgrounds. However, it was found in \cite{Golovnev:2020nln} that there are non-diagonal tetrads representing Minkowski spacetime, still consistent with the vanishing spin connection, which exhibit non-trivial dynamics. In \cite{Jimenez:2020ofm}, hints toward extra modes at 4th order around Minkowski backgrounds were found, which indicates the presence of strongly coupled fields. Cosmological perturbations around a diagonal tetrad for flat FLRW cosmology do not exhibit additional modes \cite{Golovnev:2018wbh} , and the same result is obtained when considering the spin connection \cite{Toporensky:2019mlg,Golovnev:2020aon}.
The Hamiltonian analysis suggests that more than 2 degrees of freedom propagate at the full nonlinear level \cite{Li:2011rn,Ferraro:2018tpu,Blagojevic:2020dyq}, even though their conclusions for the case $\partial_i \phi = 0$ are inconsistent (the claims are that in this case there would be two \cite{Blagojevic:2020dyq} or three of them \cite{Ferraro:2018tpu}). There is evidence that for the most general case $\partial_i \phi \neq 0$ the number of propagating degrees of freedom is 5.

For the one-parameter teleparallel gravity, which is a particular subcase of NGR, it was found in \cite{Jimenez:2019tkx} that by considering cubic interactions, the gauge symmetry in the linear theory required to avoid ghostly modes \cite{Ortin:2015hya} cannot be extended to higher orders. So far, up to our knowledge, there are no works studying perturbations for NGR around backgrounds other than Minkowski. The Hamiltonian analysis of the one-parameter teleparallel gravity has only been presented in \cite{Cheng:1988zg} (references therein present full calculations) where their claim is that the number and type of constraints depend on special values of the dynamical variables, and ``conditional bifurcations'' of the constraint algorithm appear.

We summarize the conclusions for the aforementioned theories around Minkowski, FLRW, and general backgrounds, as follows:
\begin{enumerate}
	\item Around Minkowski spacetime, TEGR recovers linearized general relativity as expected \cite{Ortin:2015hya}. The propagating degrees of freedom for $f(\mathbb{T})$ is disputed \cite{Golovnev:2020aon}. In addition to the massless spin-2 field, the one-parameter parameter family of NGR propagates a massless Kalb-Ramond field \cite{Jimenez:2019tkx,Ortin:2015hya}.
	\item Around flat FLRW spacetime, $f(\mathbb{T})$ does not seem to propagate any additional degrees of freedom, while the TEGR limit is consistent with general relativity\cite{Golovnev:2018wbh}. To our knowledge, no work has considered perturbations around FLRW backgrounds for NGR.
	\item TEGR is expected to have 2 degrees of freedom around general backgrounds, since the Hamiltonian analysis consistently gives 2 propagating degrees of freedom at the non-linear level\cite{Blagojevic:2000qs,Maluf:2001rg,Ferraro:2016wht}. To our knowledge, perturbations around general backgrounds have not been considered in the literature for teleparallel theories of gravity.
\end{enumerate}
Furthermore, the conclusion for higher order perturbations are listed as follows:
\begin{enumerate}
	\item TEGR is expected to propagate 2 degrees of freedom at all orders since the Hamiltonian analysis gives the consistent value of 2 degrees of freedom at the non-linear level\cite{Blagojevic:2000qs,Maluf:2001rg,Ferraro:2016wht}.
	\item Higher order perturbations for $f(\mathbb{T})$ around Minkowski backgrounds were considered in \cite{Golovnev:2020nln} and they found indications of extra strongly coupled modes appearing at higher order.
	\item Cubic interactions of the one-parameter family of NGR-theories around Minkowski backgrounds were considered in \cite{Jimenez:2019tkx}. They found the Kalb-Ramond field to be strongly coupled around those backgrounds.
\end{enumerate}

%
%
%
%
%----------------------------------------Outlook---------------------------------------------
%
%
%
%
%

\section{Summary and Outlook}
\label{sec:Conclusions}
We have provided a review over the current understanding of the Hamiltonian analysis in teleparallel theories of  gravity. First, we have derived the primary constraints for $f(\tngr)$ gravity, from which the derivation of the primary constraints for $f(\mathbb{T})$ and new general relativity is straightforward. Taking this theory as reference, we present a table from which the reader can compare five different classes of notation and conventions  \cite{Blagojevic:2020dyq,Blixt:2018znp,Ferraro:2018tpu,Maluf:2013gaa,Okolow:2011np}. We show that the primary constraints in these works are all consistent among them, and with the primary constraints, we derived for $f(\tngr)$. This also holds for \cite{Hohmann:2019sys,Guzman:2020kgh,Li:2011rn} which have been using similar notation as the aforementioned references. We have also provided valuable discussion on important aspects relevant for the Hamiltonian analysis in tetrad-based teleparallel theories of gravity, as for instance, the different fundamental fields considered, the choice of the spin connection, and the necessity of the ADM decomposition. We also discuss possible difficulties and misinterpretations of the Dirac-Bergmann algorithm regarding the change in the rank of the Hessian matrix and of the matrix of PB among constraints, both of them appearing for some particular subcases of modified teleparallel gravities.\\
\indent We observe that our parent theory $f(\tngr)$ possess $16+1$ canonical variables represented by the components of the tetrad $\tetrad^{A}{}_{\mu}$ and an auxiliary scalar field $\phi$. The theory is universally endowed with $4+1$ primary constraints $\momenta_A{}^{0}\approx 0$ and $\momenta_{\phi}\approx 0$, the five of them being also present in $f(\mathbb T)$ gravity. Meanwhile, for TEGR and NGR, the auxiliary scalar field $\phi$ does not appear, the theories possess only 16 canonical variables, and only the first four constraints apply. We have performed a decomposition in the remaining $12$ components of the momenta obtaining nine possible cases with different numbers of \textit{if}-constraints, in agreement with nine combinations of vanishing some of the coefficients ${}^{\mathcal I}\mathcal{A}$ (presented in Table \ref{table:NGRconstraints}). Notice that the structure of cases and their number of primary constraints is identical for both $f(\tngr)$ and NGR, but due to the appearance of $\phi$ in the primary constraints of $f(\tngr)$, it could be expected that the constraint structure of this theory is much more intricate than that of NGR. Finally, for the particular case ${}^{\mathcal{V} } \mathcal{A}=0$, ${}^{\mathcal{A} } \mathcal{A} = 0$ we obtain six primary constraints, which appear in both $f(\mathbb T)$ and TEGR, being associated with local Lorentz transformations. A good understanding of the structure of the Lorentz constraints is crucial for the comprehension of Lorentz violation in non-linear modifications of TEGR, and henceforth in the study of their viability. \\
\indent We expect that this review will pave the way for future work on the Hamiltonian analysis for teleparallel theories of gravity. A good understanding of the Hamiltonian and constraint structure of TEGR is expected if committed to the task of proposing a teleparallel-based approach to canonical quantum gravity. The outcome of the Dirac-Bergmann algorithm on the counting of degrees of freedom is also a relevant issue for $f(\mathbb T)$ gravity, on which there are disputing conclusions even for the simplest Minkowski and FLRW spacetimes. We have shown that the primary constraints for $f(\mathbb{T})$ gravity are consistent throughout the literature; therefore, the differences in the outcomes must lie in the calculations of the PB, as pointed out in \cite{Blagojevic:2020dyq}. Furthermore, the full analysis for new general relativity has only been done for the ghost-free one parameter teleparallel gravity theory in \cite{Cheng:1988zg} which does not provide many calculation details. With a better understanding for $f(\mathbb{T})$, and new general relativity (which are the most simple teleparallel theories) will give us more insights for more general modified teleparallel gravity theories. Still, Hamilton's equations have not yet been derived for any teleparallel gravity theory. And the primary Hamiltonian has not been written down without going off from the Weitzenböck gauge either. Finally, some considerations on Hessian and matrix of PB among constraints with variable rank might be necessary to take into account in this kind of models, for which the guidelines proposed in \cite{PhysRevA.54.57} can be helpful.

\section*{Acknowledgments}
The authors are grateful to R. Ferraro, A. Golovnev, J. M. Nester, and an anonymous referee for helpful discussion and reference suggestions. D. B. was supported by the University of Tartu ASTRA Project PER ASPERA, financed by the European Regional Development Fund. M. J. G. was funded by FONDECYT-ANID postdoctoral grant 3190531. M. H. and C. P. were supported by the Estonian Ministry for Education and Science through the Personal Research Funding Grants PRG356 and PSG489. D. B., M. H. and C. P. were supported by the European Regional Development Fund through the Center of Excellence TK133 ``The Dark Side of the Universe''.

	\bibliographystyle{unsrt}
	\bibliography{HamiltonianDictionaryArxivFinal}

\end{document}